\documentclass[pra,reprint,twocolumn,showpacs,superscriptaddress]{revtex4-1}
\usepackage{amsmath,amssymb,mathrsfs}
\usepackage{psfrag}
\usepackage{graphicx}
\usepackage{graphics}
\usepackage{epsfig}
\usepackage{bm}
\usepackage{color}
\usepackage{verbatim,color,ulem}
\usepackage{physics}
\usepackage[dvipdfmx,colorlinks=true,linkcolor=blue,urlcolor=blue,
citecolor=blue]{hyperref}

\newcommand{\beq}{\begin{equation}}
\newcommand{\eeq}{\end{equation}}
\newcommand{\bal}{\begin{aligned}[b]}
\newcommand{\eal}{\end{aligned}}
\newcommand{\bseq}{\begin{subequations}}
\newcommand{\eseq}{\end{subequations}}
\newcommand{\beqa}{\begin{eqnarray}}
\newcommand{\eeqa}{\end{eqnarray}}
\newcommand{\cred}{\color{black}}

\begin{document}

\title{Amplitude, phase, and topological fluctuations shaping the complex phase diagram of two-dimensional superconductors}

\author{Koichiro Furutani}
\email{koichiro.furutani@phd.unipd.it}
\affiliation{Department of Applied Physics, Nagoya University, Nagoya 464-8603, Japan}
\affiliation{Institute for Advanced Research, Nagoya University, Nagoya 464-8601, Japan}
\author{Giovanni Midei}
\affiliation{School of Science and Technology, Physics Division, 
Universit\`a di Camerino, Via Madonna delle Carceri 9, 62032 Camerino, Italy}
\author{Andrea Perali}
\affiliation{School of Pharmacy, Physics Unit, 
Universit\`a di Camerino, Via Madonna delle Carceri 9, 62032 Camerino, Italy}
\author{Luca Salasnich}
\affiliation{Dipartimento di Fisica e Astronomia 'Galileo Galilei'  
and QTech Center, Universit\`a di Padova, via Marzolo 8, 35131 Padova, Italy}
\affiliation{Istituto Nazionale di Fisica Nucleare, Sezione di Padova, 
via Marzolo 8, 35131 Padova, Italy}
\affiliation{Istituto Nazionale di Ottica del Consiglio Nazionale delle Ricerche, 
via Carrara 2, 50019 Sesto Fiorentino, Italy}

\date{\today}

\begin{abstract}
We study the amplitude and phase fluctuations of the Ginzburg-Landau quasiorder 
parameter for superconductors in two spatial dimensions. 
Starting from the mean-field critical temperature $T_{\mathrm{c}0}$, we calculate the beyond-mean-field critical temperature $T_{\rm c}$ by including thermal fluctuations of the quasiorder parameter within the Gaussian level. 
Moreover, from our beyond-mean-field results, we derive the Berezinskii-Kosterlitz-Thouless critical temperature $T_{\rm BKT}$, which takes into account topological vortex-antivortex excitations in the phase fluctuations as well as the amplitude fluctuations, to obtain the shifts of transition temperatures. 
We elucidate how the Gaussian thermal fluctuations and phase fluctuations associated with vortex excitations affect thermodynamic properties by determining the $H$-$T$ phase diagram for a type-II superconductor and computing the critical behaviors of the heat capacity, which are experimentally accessible, allowing the characterization of the cascade of different kinds of fluctuations in 2D superconductors. 
\end{abstract}


\maketitle

\section{Introduction}

Phase transitions in low-dimensional systems are central targets in modern physics and critical fluctuations play a crucial role in shaping their complex phase diagram, while featuring several anomalous properties detectable in experiments. 
The Ginzburg-Landau theory describes the behavior of an order parameter in a superconductor near the superconducting transition under spontaneous symmetry breaking, which captures a broad range of phase transitions including second-order phase transitions of superconductors or superfluids \cite{ginzburg}, multiband superconductors \cite{milosevic2010,speight2011,morimoto2022,milosevic2022}, and twisted bilayer graphene \cite{law2024}. 
Two-dimensional superconductors or superfluids, however, undergo a topological phase transition without spontaneous symmetry breaking, which is referred to as the Berezinskii-Kosterlitz-Thouless (BKT) transition \cite{kosterlitz,nelson}. 
BKT transitions were first observed in a thin $^{4}\mathrm{He}$ film \cite{bishop78}. 
Later, it was experimentally observed also in thin and disordered superconducting films \cite{mondal2011,yong13,iwasa2021,weitzel2023} and a $^{39}\mathrm{K}$ atomic gas \cite{hadzibabic2021} through the measurement of sound velocities \cite{furutani2021,salasnich2022}. 
Phase fluctuations associated with the vortex excitations suppressing the superconducting transition temperature play a key role in the BKT transition. 
As exemplified by BKT physics, identifying what kind of fluctuations govern the microscopic behavior of a material is imperative to reveal the phase diagram. 
For instance, it has been investigated through measurements of thermodynamic quantities such as the heat capacity \cite{wang2001,taylor2007,nguyen2020,shibauchi2023} and the resistivity or the $I$-$V$ characteristics \cite{iwasa2021}. 
The recent experiment in Ref.~\cite{shibauchi2023} reports spectroscopic and thermodynamic investigations of the iron-based superconductor $\mathrm{FeSe}_{1-x}\mathrm{S}_{x}$ throughout the Bardeen-Cooper-Schrieffer-Bose-Einstein-condensate (BCS-BEC) crossover \cite{randeria1993}, suggesting the dominant role of Gaussian thermal fluctuations in the vicinity of criticality. 

In two-dimensional materials, in particular, the occurrence of off-diagonal long-range order (ODLRO) is ruled out at finite temperatures \cite{merminwagner1966,hohenberg1967}, and the BKT transition is an infinite-order phase transition toward a quasi-long range ordered state \cite{altlandsimons}, which hinders singularities of thermodynamic quantities in the vicinity of the transition point making the role of respective fluctuations difficult to establish across the BKT transition. 
These specific features of two-dimensional materials have motivated several attempts to clarify the dominant fluctuations and their evolution from the normal to the superconducting state both theoretically and experimentally. 
In the normal phase of cuprate superconductors in the underdoped regime, where the electronic behavior is practically two-dimensional, phase fluctuations are predicted to mainly contribute to the quasiparticle spectra \cite{franz1998}. 
The measurement of the Nernst coefficient supports that vortex excitations survive even well above the critical temperature, but quite below the pseudogap temperature; the doping dependence of the vortex excitations onset temperature shows the same dome-like behavior of the critical temperature \cite{ong2006}. 
The specific heat of cuprates is also affected by the pseudogap and exhibits similar peaked behavior as in $\mathrm{FeSe}$ \cite{loram2001}. 
Therefore, in two-dimensional strongly coupled superconductors, the superconducting transition is approached from the normal state through a weak-pseudogap phase at high temperature, where only amplitude fluctuations (Gaussian-like) are present, and then through a strong-pseudogap phase close to critical temperature, where also phase fluctuations are important and add their effects to the amplitude fluctuations, leading to well-formed pseudogap features and long-lived Cooper pairs (a quasicondensate configuration). 
In a two-dimensional Fermi atomic gas above the superfluid transition temperature, a $T$-matrix analysis revealed that pseudogap grows with strong pairing fluctuations across the BCS-BEC crossover \cite{strinati2015}, which is in fair agreement with the experimental measurements of photoemission spectra with $^{40}\mathrm{K}$ atoms \cite{feld2011,frohlich2012}. 
Theoretical works also predict the nonmonotonic behavior of the BKT transition temperature in terms of the pairing interaction strength \cite{levin2020,wang2020} and the upper bound \cite{shi2023} across the BCS-BEC crossover. 
However, in the context of ultracold atoms, experimental or theoretical
investigations of the cooperative interplay between phase and amplitude fluctuations in determining strong pseudogap features are lacking, also considering that $T$-matrix approaches in two dimensions close to the critical temperature are ill-defined. 
The above discussion motivates us to investigate the cascade of fluctuations in two-dimensional superconductors exploring a wide temperature range across the critical temperature to disentangle the effect on the system properties of each
type of fluctuation. 

In this work, we consider two-dimensional superconductors to analyze the effects of both the Gaussian thermal fluctuations and topological phase fluctuations, which result in the BKT transition within the Ginzburg-Landau theory. 
We examine the fluctuations of different natures characterizing the phase diagram of the system and discuss their roles in the renormalization of the superconducting critical temperature.  
In Sec.~\ref{SecGL}, we start from the Ginzburg-Landau free energy and show the mean-field analysis. 
In Sec.~\ref{SecGaussian}, we include thermal fluctuations at the Gaussian level, which lowers the critical temperature from the mean-field one. 
In Sec.~\ref{SecBKTfl}, we further incorporate the vortex excitations responsible for the BKT transition by solving the Nelson-Kosterlitz (NK) renormalization group equations. 
We discuss the roles of both the Gaussian thermal fluctuations and phase fluctuations associated with vortex excitations, and obtain a formula for the shift of critical temperatures. 
To this end, we employ a semi-analytic relation between the bare and the renormalized phase stiffness. 
Finally, we determine and discuss the $H$-$T$ phase diagram of a typical type-II superconductor in Sec.~\ref{SecHc} and the critical behavior of the heat capacity in Sec.~\ref{Secheat}, which highlights the roles of fluctuations. 
In particular, the phase fluctuations associated with vortex excitations turn out to drastically change the critical behavior of the heat capacity. 
Our results suggest that the measurement of the heat capacity in the vicinity of critical temperature can be utilized to verify what types of fluctuations govern a two-dimensional superconducting material. 
Section \ref{Secconclusion} summarizes our results. 
Details of numerical and analytic computations are reported in Appendices \ref{Appeta}, \ref{AppCv}, and \ref{AppCv2d}.

\section{Hohenberg-Mermin-Wagner's theorem and BKT transition}
Before moving on to our analysis, we review quasi-long-range order (qLRO) in two-dimensional superconductivity and briefly summarize how to investigate the effects of amplitude fluctuations in two-dimensional superconductors subject to the BKT transition. 

Hohenberg-Mermin-Wagner's theorem rules out spontaneous symmetry breaking at finite temperatures in an infinite-size system hosting short-range interaction and a continuous symmetry with spatial dimensions lower than three \cite{merminwagner1966,hohenberg1967}. 
It prohibits the occurrence of ODLRO at finite temperatures in two dimensions. 
With a superconducting order parameter $\psi$, it states the vanishing ensemble average $\langle\psi\rangle=0$. 
This originates from enhanced infrared fluctuations in lower dimensions that break the ordered state in an infinite-size system. 
However, Berezinskii, Kosterlitz, and Thouless pointed out that dissipationless flow is allowed in two dimensions by the formation of vortex-antivortex pairs at low temperatures even though the ODLRO is absent \cite{kosterlitz,nelson}. 
This is due to the fact that topological phase fluctuations of the order parameter can only occur in the form of vortex-antivortex pairs below the BKT transition temperature since configurations including free vortices are energetically forbidden. 
Unlike the free vortex case, a pair of vortices with opposite winding numbers leads to far-field configurations that can be continuously distorted to the uniform ground state. 
In this way, the detrimental impact of the topological defects on the phase coherence gets dramatically suppressed. 
The presence of these topological defects and the interaction among them leave very peculiar signatures in the normal to superconducting BKT transition and in the temperature dependence of different physical observables, as investigated in our work. 
Clearly, the Cooper pairing setting at higher temperatures is the other key feature for superconductivity, being the building block of a finite amplitude of the superconducting wavefunction without which the phase itself, fluctuating or ordered, cannot exist. 
As the temperature rises, the dissociation of neutral vortex-antivortex pairs occurs at a transition temperature and then the superconductivity is broken, which is the mechanism of the BKT transition. 
Then, the vortex fluctuations results in $\langle\abs{\psi}\rangle\neq 0$ in the low-temperature phase even though $\langle\psi\rangle=0$ \cite{larkin}. 
We may call $\psi$ the {\it quasi}order parameter.  
The square of the ensemble average of the quasiorder parameter {\it modulus} can be indeed interpreted as the superfluid density. 
A salient feature of the BKT superconductor is the qLRO in the low-temperature phase described by the power-law decay of the single-body density matrix $\langle\psi({\bf r})\psi({\bf r'})\rangle\propto \abs{{\bf r}-{\bf r'}}^{-\eta}$. 
The algebraic exponent $\eta$ is related to the superfluid density and is a good indicator to verify the BKT transition. 

There are several attempts to investigate the effects of amplitude fluctuations on the two-dimensional superconductors. 
One of the major approaches is the $T$-matrix analysis. 
The $T$-matrix analysis is, however, ill-defined due to the infrared divergence in two dimensions as dictated by Hohenberg-Mermin-Wagner's theorem. 
This problem can be solved by exploiting an infrared cutoff $k_{0}$ to keep the infrared divergence under control \cite{strinati2015}. 
In this way, in Ref.~\cite{strinati2015}, a successful comparison between theory and experiments for a 2D trapped Fermi system has been reported in the intermediate temperature range between the low-temperature region approaching the critical temperature from above and the high-temperature region where the $T$-matrix describes a semiclassical system of attractive fermions. 
Once fluctuations are taken into account in this manner, superconducting fluctuations can be quantitatively evaluated by the $T$-matrix approach or the Ginzburg-Landau approach, which yields almost the same results. 
This treatment is also supported by experimental reports of the BCS-Bose-Einstein-quasicondensation (BEqC) crossover phenomena including pseudogap formation and pair-size shrinking in 2D superconductors of $\mathrm{Li}_{x}\mathrm{ZrNCl}$ \cite{iwasa2021}, which are similar to 3D cases. 
It is because the phenomenology is local in character and connected with a short-range scale common with 3D cases except in the vicinity of the BKT transition. 
In this paper, we investigate effects of fluctuations on the BKT transition by following this strategy.

\section{Ginzburg-Landau functional}\label{SecGL}

Close to the critical temperature the free energy of a single-band 
superconducting material can be written as 
\beq 
F = F_{\rm n} + F_{\rm s} \; , 
\eeq
where $F_{\rm n}$ is the contribution due to the normal component and $F_{\rm s}$ is the contribution due to the emergence of a superconducting quasiorder parameter $\psi({\bf r})$ below the critical temperature. 
Within the Ginzburg-Landau approach \cite{ginzburg}, for a two-dimensional system of area $L^2$, the super component $F_{\rm s}$ is given by 
\beq 
F_{\rm s} = \int_{L^2}  \dd[2]{\bf r}
\left\{ a(T) \, |\psi({\bf r})|^2 + {b\over 2} \, 
|\psi({\bf r})|^4 + \gamma \, | \grad \psi({\bf r})|^2 \right\} \; , 
\label{gl}
\eeq
where 
\beq 
a(T) = \alpha \, k_{\rm B} \, (T - T_{\mathrm{c}0})  
\label{chiea}
\eeq
is a parameter which depends on the temperature $T$ and becomes zero at the mean-field (MF) critical temperature $T_{\mathrm{c}0}$, while $b>0$ and $\gamma>0$ are temperature-independent parameters with $k_{\rm B}$ being the Boltzmann constant. 
The energy functional \eqref{gl} and the values of the parameters $\alpha$, $b$ and $\gamma$ can be deduced from the microscopic BCS theory as \cite{bcs,gorkov}
\beq
\alpha=\frac{4\pi^{2}}{7\zeta(3)}\frac{T_{\mathrm{c}0}}{T_{\rm F}}, \quad
b=\frac{7\zeta(3)}{8\pi^{2}}\frac{\alpha^{2}}{\nu}, \quad
\gamma=\frac{\hbar^{2}}{4m},
\label{bcsparameter}
\eeq 
with $T_{\rm F}$ being the Fermi temperature, $m$ being the electronic mass, $\nu=m/\pi\hbar^{2}$ being the density of states for two-dimensional electrons having a parabolic energy dispersion, and $\zeta(x)$ being the zeta function. 

The partition function ${\cal Z}$ of the system is given by 
\beq 
{\cal Z} =  e^{-\beta F_{\rm n}} \ {\cal Z}_{\rm s} \; , 
\eeq
where 
\beq 
{\cal Z}_{\rm s} = \int \mathcal{D}[\psi({\bf r})] \, e^{-\beta F_{\rm s}[\psi({\bf r})]} \;  
\eeq
is the partition function of the superconducting component 
with $\beta = 1/k_{\rm B}T$. The thermal average of an observable $\mathcal{O}$ 
that is a functional of $\psi({\bf r})$ reads 
\beq 
\langle \mathcal{O} \rangle = 
{1\over {\cal Z}_{\rm s}} \int \mathcal{D}[\psi({\bf r})] \, \mathcal{O}[\psi({\bf r})] 
\, e^{-\beta F_{\rm s}[\psi({\bf r})]}  \; . 
\eeq

\begin{figure}[t]
\centering
\includegraphics[width=80mm]{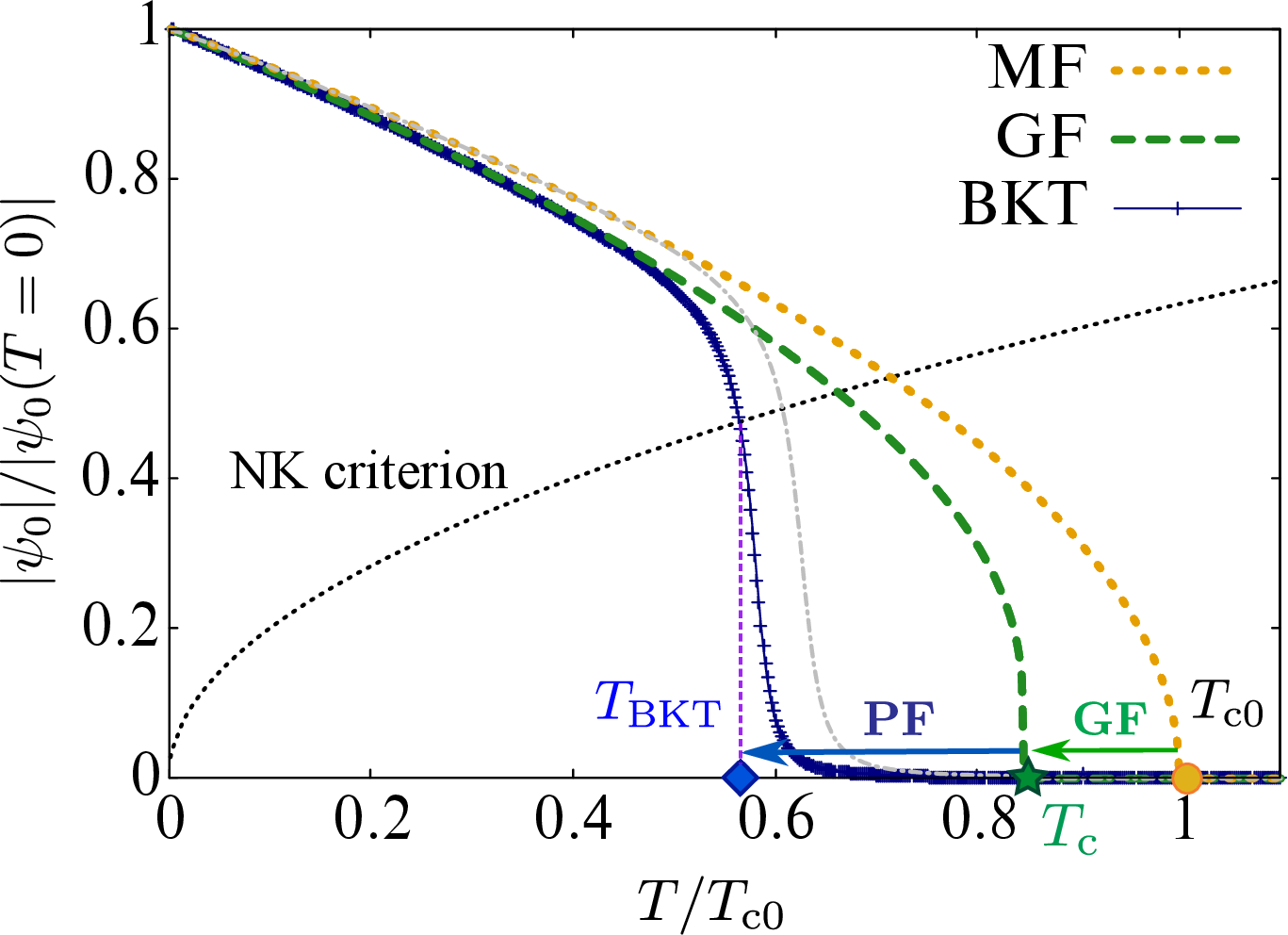}
\caption{Modulus of the quasiorder parameters as a function of the temperature $T$. 
The quasiorder parameter $\psi_{0}$ is obtained at the MF level with Eq.~\eqref{psi0} (dotted curve), $\psi_{0}^{({\rm GF})}$ is derived including the Gaussian fluctuations (GF) in Eq.~\eqref{phi0} with Eq.~\eqref{eta2sf} (dashed curve), and $\psi_{\rm R}^{({\rm BKT})}=\sqrt{J_{\rm R}(T)/2\gamma}$ is determined by Eqs.~\eqref{rg1} and \eqref{rg2} including the phase fluctuations (PF) associated with vortex excitations responsible for the BKT transition under $l_{\rm max}=\ln(\pi/\xi k_{0})$ (solid curve). 
We set $\mathrm{Gi}=0.1$. 
The gray dotted-dashed curve represents the order parameter obtained by Eqs.~\eqref{rg1} and \eqref{rg2} under the MF bare values without the Gaussian fluctuations. 
These quasiorder parameters vanish at different critical temperatures: $T_{\mathrm{c}0}$ at the MF level, $T_{\rm c}$ at the GF level without vortices, and $T_{\rm BKT}$ including the vortex excitations. 
The thin dotted black curve represents $k_{\rm B}T=\pi\gamma\abs{\psi_{0}}^{2}$, which stands for the NK criterion. 
It intersects with the quasiorder parameter at $T_{\rm BKT}$ (vertical violet dashed line). }
\label{FigTc0TcTBKT}
\end{figure}

By assuming a real and uniform quasiorder parameter, i.e., 
\beq 
\psi({\bf r}) = \psi_0 \; , 
\label{simple}
\eeq
the energy functional \eqref{gl} with Eqs.~\eqref{chiea} and \eqref{simple} 
becomes 
\beq 
{F_{\mathrm{s}0}[\psi_{0}]\over L^2} = a(T) \, \psi_0^2 + 
{b\over 2} \, \psi_0^4  \; . 
\label{uniform}
\eeq
Minimizing $F_{\mathrm{s}0}$ with respect to $\psi_0$, one immediately finds 
\beq 
a(T) \, \psi_0 + b \, \psi_0^3 = 0 \; , 
\label{pio}
\eeq
and consequently 
\beq 
\psi_0 =  \left\{ \begin{array}{ll}
          0 & \mbox{ for } T \geq T_{\mathrm{c}0}, \\\\ 
          \displaystyle\sqrt{-{a(T) \over b}} & \mbox{ for } T < T_{\mathrm{c}0}. 
\end{array} \right. \;   
\label{psi0}
\eeq
The uniform quasiorder parameter $\psi_0$ becomes different from zero 
only below the MF critical temperature $T_{\mathrm{c}0}$, 
where, by definition, $T_{\mathrm{c}0}$ is such that 
\beq 
a(T_{\mathrm{c}0}) = 0 \; . 
\eeq
The temperature dependence of the mean-field quasiorder parameter \eqref{psi0} is shown 
in Fig.~\ref{FigTc0TcTBKT}.

\section{Gaussian fluctuations of the quasi-order parameter}\label{SecGaussian}

Extremizing the functional \eqref{gl} with respect to $\psi^*({\bf r})$ 
one gets the Euler-Lagrange equation 
\beq 
a(T) \, \psi + b \, |\psi|^2\psi - \gamma \, \grad^2 \psi = 0 \; . 
\label{gle}
\eeq
We write the space-dependent quasiorder parameter $\psi({\bf r})$ {\cred as
\beq
\bal 
\psi({\bf r}) &=[\psi_{0}^{({\rm GF})}+\sigma({\bf r})]e^{i\theta_{0}({\bf r})} \\
&\simeq \psi_0^{({\rm GF})} + \eta({\bf r}) \; , 
\eal
\label{shift}
\eeq 
where $\eta({\bf r})=\sigma({\bf r})+i\theta_{0}({\bf r})\psi_{0}^{({\rm GF})}$ represents fluctuations of both amplitude and phase to the first order} with respect to the real and uniform configuration $\psi_0^{({\rm GF})}$ with the condition 
\beq 
\langle \eta \rangle = \langle \eta^* \rangle = 0 \; , 
\eeq
where $\langle \cdot \rangle$ is the thermal average. 
Inserting Eq.~\eqref{shift} into Eq.~\eqref{gle}, we find 
\beqa 
a(T) \, \psi_0^{({\rm GF})} &+& b \, (\psi_0^{({\rm GF})})^3  + 
a(T) \, \eta + 2 b \, (\psi_0^{({\rm GF})})^2 \eta 
\nonumber 
\\
&+& b \, (\psi_0^{({\rm GF})})^2 \eta^* + 
2 b \, \psi_0^{({\rm GF})} |\eta|^2 
\nonumber 
\\
&+& b \, \psi_0^{({\rm GF})} \eta^2 + b \, |\eta|^2 \eta - 
\gamma \grad^2 \eta = 0 \; , 
\label{completa}
\eeqa
and after thermal averaging we obtain 
\beq 
\left[ a(T)+2 b \, \langle |\eta|^2 \rangle 
+ b \, \langle \eta^2 \rangle \right] \, \psi_0^{({\rm GF})} + 
b \, (\psi_0^{({\rm GF})})^3 +  b \, \langle |\eta|^2\eta \rangle = 0 \; .  
\label{piopio}
\eeq
Clearly, only by removing all thermal averages, one recover Eq.~\eqref{pio} and find that $\psi_0^{({\rm GF})}$ is equal to $\psi_0$ and given by Eq.~\eqref{psi0}. 
In the framework of the Gross-Pitaevskii equation, this is called the Bogoliubov approximation \cite{zaremba}. 

We take into account thermal fluctuations of the quasiorder parameter keeping $\langle |\eta|^2 \rangle$ but neglecting the anomalous averages $\langle \eta^2 \rangle$ and $\langle |\eta|^2\eta \rangle$. 
This is the so-called Popov approximation \cite{zaremba}. 
In this way, we get 
\beq 
\left[ a(T)+2 b \, \langle |\eta|^2 \rangle \right] 
\, \psi_0^{({\rm GF})} + b \, (\psi_0^{({\rm GF})})^3 = 0 \; , 
\label{palle}
\eeq
and consequently 
\beq 
\psi_0^{({\rm GF})} =  \left\{ \begin{array}{ll}
          0 & \mbox{ for } T \geq T_{c}, \\\\
          \displaystyle\sqrt{-{a(T)+2b \langle |\eta|^2 \rangle \over b}} 
& \mbox{ for } T < T_{c}. 
\end{array} \right. \;  
\label{phi0}
\eeq
In this case, the uniform quasiorder parameter $\psi_0^{({\rm GF})}$ becomes 
different from zero only below the beyond-MF critical temperature $T_{\rm c}$ determined by 
\beq 
a(T_{\rm c}) + 2 b \, \langle |\eta|^2 \rangle_{\rm c} = 0 \; , 
\label{cicredi}
\eeq
with $\langle\abs{\eta}^{2}\rangle_{\rm c}\equiv \langle\abs{\eta}^{2}\rangle_{T\to T_{\rm c}^{+}}$. 
The free energy generating Eq.~\eqref{palle} including the Gaussian thermal fluctuations is given by
\beq
\frac{F_{\mathrm{s}0}^{(\text{GF})}}{L^{2}}=\left[a(T)+2b\langle\abs{\eta}^{2}\rangle\right](\psi_{0}^{(\text{GF})})^{2}+\frac{b}{2}(\psi_{0}^{(\text{GF})})^{4}. 
\label{uniform-gf}
\eeq
It is important to stress that, to treat self-consistently the fluctuating field $\eta({\bf r})$, the Popov approximation requires $|\eta|^2 \eta \simeq 2 \langle |\eta|^2\rangle \eta$, $|\eta|^2 \simeq \langle |\eta|^2 \rangle$, and $\eta^2 \simeq 0$ in Eq.~\eqref{completa} \cite{zaremba}, which becomes 
\beq 
a(T) \, \eta + 2 b \, (\psi_0^{({\rm GF})})^2 \eta + 
2 b \, \langle |\eta|^2 \rangle \eta 
+ b \, (\psi_0^{({\rm GF})})^{2} \eta^* - \gamma \grad^2 \eta = 0 \; , 
\label{quasicompleta}
\eeq
taking into account Eq.~\eqref{palle}. Notice that the Popov approximation 
gives a gapless spectrum, as required by the Goldstone theorem \cite{zaremba}. 

Let us work with $T\geq T_{\rm c}$ where $\psi_0^{({\rm GF})}=0$. 
Then, Eq.~\eqref{quasicompleta} reduces to 
\beqa 
a(T) \, \eta + 2 b \, \langle |\eta|^2 \rangle \eta 
- \gamma \grad^2 \eta = 0 \; , 
\label{etaeqnormal}
\eeqa
which is the Euler-Lagrange equation of the Gaussian energy functional 
\beq
\bal
F_{\eta}^{\rm Popov}(T\ge T_{\rm c}) &= \int_{L^2}  \dd[2]{\bf r}
\Big[ (a(T)|\eta({\bf r})|^2 + 2 b \langle |\eta|^2 \rangle ) 
|\eta({\bf r})|^2 \\
&+ \gamma \, | \grad \eta({\bf r})|^2 \Big] \; .  
\eal
\eeq
Expanding the field $\eta({\bf r})$ in a plane-wave basis, i.e., 
\beq 
\eta({\bf r}) = \frac{1}{L}\sum_{\bf k} {\tilde \eta}_{\bf k} e^{i {\bf k}\cdot 
{\bf r}} \; , 
\eeq 
it is straightforward to find 
\beq 
F_{\eta}^{\rm Popov}(T\ge T_{\rm c}) = \sum_{\bf k} \left[ a(T) + 2 b \langle |\eta|^2 \rangle + 
\gamma k^2 \right] |{\tilde \eta}_{\bf k}|^2  \; . 
\label{Fetanormal}
\eeq
The partition function ${\cal Z}$ of the system is then given by 
\beq 
\bal
{\cal Z} &=  e^{-\beta F_{\rm n}} \ \int \mathcal{D}[\eta({\bf r})] \, 
e^{-\beta F_{\eta}^{\rm Popov}[\eta({\bf r})]} \\
&=e^{-\beta F_{\rm n}-\beta F_{\rm fl}^{+}} \; , 
\eal
\label{superz}
\eeq
where 
\beq
\bal
F_{\rm fl}^{+}&=-\frac{1}{\beta}\ln{\int \mathcal{D}[\eta({\bf r})] \, 
e^{-\beta F_{\eta}^{\rm Popov}[\eta({\bf r})]}} \\
&=-\frac{1}{\beta}\sum_{\bf k}\ln{\left(\dfrac{\pi k_{\rm B}T}{a(T)+2b\langle\abs{\eta}^{2}\rangle+\gamma k^{2}}\right)},
\eal
\label{Fflabove}
\eeq
represents the contribution of the thermal fluctuations to the free energy above $T_{\rm c}$.
The thermal average 
\beq 
\langle |\eta|^2 \rangle = 
{1\over {\cal Z}_{\eta}} \int \mathcal{D}[\eta({\bf r})] \, |\eta({\bf r})|^2  
\, e^{-\beta F_{\eta}^{\rm Popov}[\eta({\bf r})]}  
\eeq
reads 
\beq 
\langle |\eta|^2 \rangle = {1\over \beta L^2} 
\sum_{\bf k} {1 \over a(T) + 2 b \langle |\eta|^2 \rangle 
+ \gamma k^2} \; , 
\label{eta2}
\eeq 
with $\mathcal{Z}_{\eta}\equiv \int\mathcal{D}[\eta({\bf r})]\,\mathrm{exp}[-\beta F_{\eta}^{\rm Popov}[\eta({\bf r})]]$. 
At $T\to T_{\rm c}+$, where Eq.~\eqref{cicredi} holds, we get 
\beq 
\langle |\eta|^2 \rangle_{\rm c} = {1\over L^2} 
\sum_{\bf k} {k_{\rm B}T_{\rm c} \over \gamma k^2} \; . 
\eeq
In the continuum of momenta ${\bf k}$, where $\sum_{\bf k}=L^2 \int \dd[2]{\bf k}/(2\pi)^2$, we assume that some physical cutoff constrains the values of $k$ \cite{larkin}. 
Under this assumption, we find 
\beq 
\langle |\eta|^2 \rangle_{\rm c} = {k_{\rm B}T_{\rm c} \over 2\pi \gamma} 
\ln\left({\Lambda\over k_0}\right) \; , 
\label{eta2c} 
\eeq
where $k_0$ is an infrared cutoff and $\Lambda$ an ultraviolet cutoff. 
Then, Eq.~\eqref{cicredi} gives 
\beq 
{T_{\mathrm{c}0} - T_{\rm c} \over T_{\rm c}} = 4 \, \mathrm{Gi} \, 
\ln\left({\Lambda\over k_0}\right),
\label{Tc0Tcdiff}
\eeq
with 
\beq 
\mathrm{Gi} = {b\over 4\pi \alpha \gamma}
\label{levanyuk}
\eeq
the Ginzburg-Levanyuk number. 
With Eqs.~\eqref{bcsparameter}, Eq.~\eqref{levanyuk} reduces to $\mathrm{Gi}=\pi T_{\mathrm{c}0}/2T_{\rm F}$. 
According to Larkin and Varlamov \cite{larkin}, one has 
{\cred 
\beq
\Lambda = \frac{1}{2\xi}, \quad\quad k_0=\frac{\sqrt{\mathrm{Gi}}}{\xi},
\label{Lambdak0}
\eeq
with $\xi=(\gamma/\alpha k_{\rm B}T_{\rm c})^{1/2}$ the coherence length. }
It follows that ${\Lambda/k_0}=1/(4\, \mathrm{Gi})^{1/2}$ and consequently 
\beq
{T_{\mathrm{c}0} - T_{\rm c} \over T_{\rm c}} = 2 \, \mathrm{Gi} \, 
\ln\left(\dfrac{1}{4\mathrm{Gi}}\right) \; . 
\label{shift2d}
\eeq

The shift in Eq.~\eqref{shift2d} coincides with Eq.~(2.120) in Ref.~\cite{larkin} obtained via renormalization group analysis of the Ginzburg-Landau coefficients, which justifies the introduction of the infrared momentum cutoff $k_{0}$ in Eqs.~\eqref{Lambdak0}. 
We assume a sufficiently large system size so that $k_{0}$ is larger than the infrared momentum cutoff determined by the system size $k_{\rm min}=\pi/L$. 
Then, $k_{0}$ practically plays the role of the infrared momentum cutoff fulfilling the consistency with the renormalization of the Ginzburg-Landau coefficients. 
The valid domain of the Ginzburg-Landau analysis is $\mathrm{Gi}\ll1$. 
It makes $k_{0}$ progressively smaller so does  $k_{\rm min}$ to keep the inequality $k_{0}>k_{\rm min}$, which is realized by scaling to infinity the system size. 
Interestingly, our choice of the infrared cutoff can be written as $k_{0}=(4\mathrm{Gi})^{1/2}\,\Lambda$ in such a way as to make clear its behavior approaching the weak-coupling regime. 
When the ultraviolet cutoff $\Lambda$ is fixed and $\mathrm{Gi}$ goes to zero in the large density regime, $k_{0}$ scales toward zero, enhancing the region of validity of our approach. 
For very small $\mathrm{Gi}$, the mean-field and the beyond mean-field Gaussian fluctuations physics will survive also very close to the phase transition. 
This justifies our choice of $k_{0}$, having the expected behavior when the weak-coupling BCS regime is approached by tuning the density (see the discussion reported in Ref.~\cite{strinati2015}). 
Note that this result takes into account both amplitude and phase fluctuations of the quasiorder parameter at the Gaussian level. 
See also Refs.~\cite{larkin,sala2019,sala2020}. 
However, Eq.~\eqref{shift2d} does not include the effects of the topological defects (quantized vortices and antivortices).

\begin{figure}[t]
\centering
\includegraphics[width=80mm]{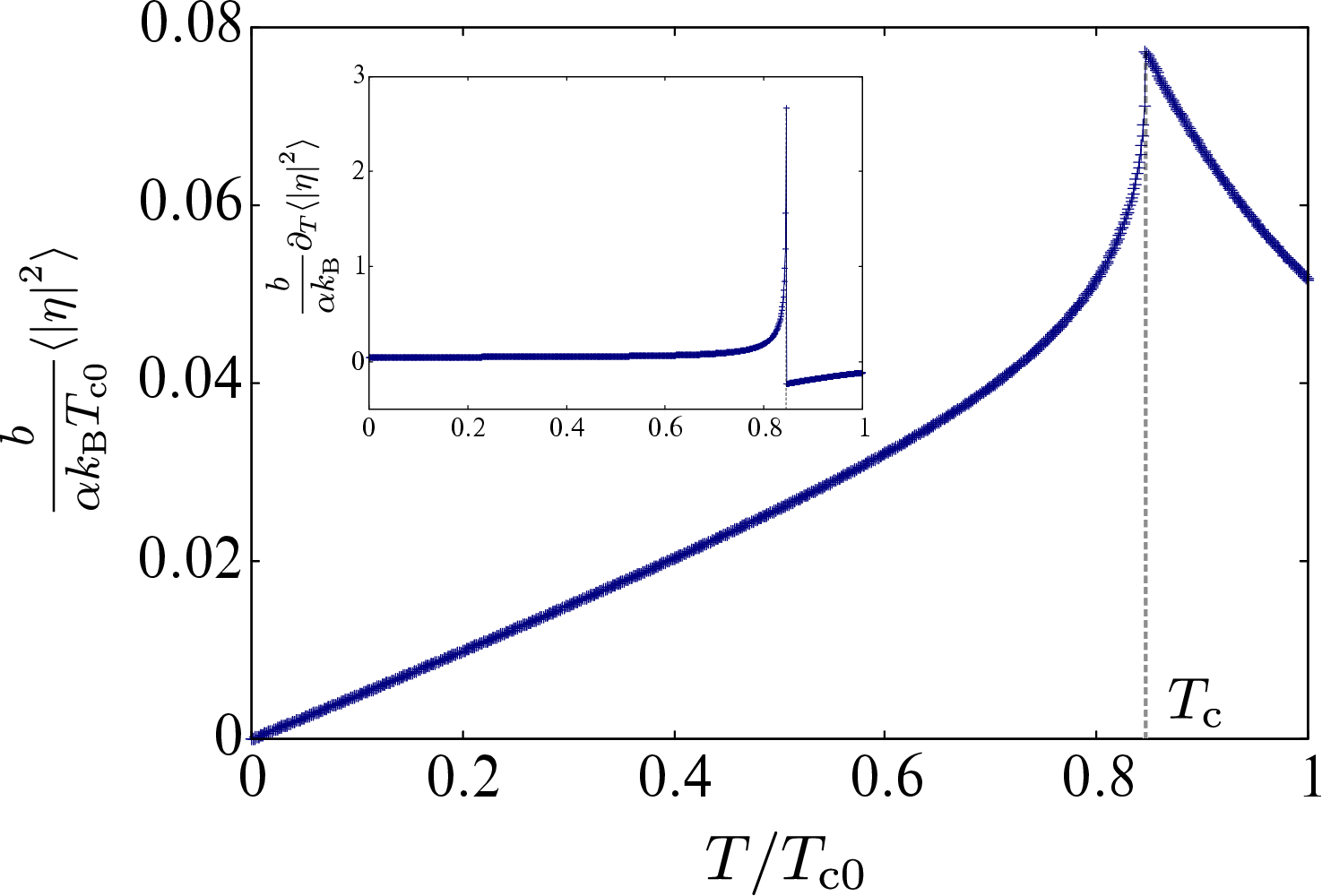}
\caption{Magnitude of the correction to the quasiorder parameter by the 
Gaussian fluctuations $\langle|{\eta}|^{2}\rangle$ calculated 
from Eq.~\eqref{eta2sf} below $T_{\rm c}$ and Eq.~\eqref{eta2} 
above $T_{\rm c}$ with $\mathrm{Gi}=0.1$. 
The inset shows the temperature derivative $\partial_{T}\langle|{\eta}|^{2}\rangle(T)$. }
\label{Figeta2}
\end{figure}

\subsection{Gaussian fluctuations below $T_{\rm c}$}

To compute Eq.~\eqref{phi0}, we examine the Gaussian fluctuations (GF) below $T_{\rm c}$. 
Instead of Eq.~\eqref{etaeqnormal}, $\langle|\eta|^{2}\rangle$ is determined by Eq.~\eqref{quasicompleta}. Then, Eq.~\eqref{quasicompleta} reads
\beq
\dfrac{\delta F_{\eta}^{\rm Popov}}{\delta \eta^{*}}=0,
\eeq
with
\beq
\bal
&F_{\eta}^{\rm Popov}\equiv \int \dd[2]{\bf r}\Bigg[\left[a(T)
+2b(\psi_{0}^{({\rm GF})})^{2}\right]|{\eta({\bf r})}|^{2} \\
&+2b\langle|{\eta}|^{2}\rangle|{\eta({\bf r})}|^{2} 
+\dfrac{b}{2}(\psi_{0}^{({\rm GF})})^{2}\left[\eta({\bf r})^{2}
+\eta^{*}({\bf r})^{2}\right] \\
&+\gamma|{\grad\eta({\bf r})}|^{2}\Bigg] \\
&=\sum_{\bf k}\Big[\left(a(T)+2b\langle|{\eta}|^{2}\rangle
+3b(\psi_{0}^{({\rm GF})})^{2}+\gamma k^{2}\right)\Tilde{\eta}_{\bf k}'^{2} \\
&+\left(a(T)+2b\langle|{\eta}|^{2}\rangle
+b(\psi_{0}^{(\mathrm{GF})})^{2}+\gamma k^{2}\right)\Tilde{\eta}_{\bf k}''^{2}\Big], 
\eal
\label{Fetapopov}
\eeq
where $\Tilde{\eta}_{\bf k}'$ and $\Tilde{\eta}_{\bf k}''$ are the real part 
and imaginary part, respectively, of $\Tilde{\eta}_{\bf k}=
\Tilde{\eta}_{\bf k}'+i\Tilde{\eta}_{\bf k}''$. 
In Eq.~\eqref{Fetapopov}, the first term proportional to $\Tilde{\eta}_{\bf k}'^{2}$ corresponds to the amplitude fluctuations while the second term proportional to $\Tilde{\eta}_{\bf k}''^{2}$ corresponds to the phase fluctuations \cite{larkin}. 
By setting $\psi_{0}^{({\rm GF})}=0$, Eq.~\eqref{Fetapopov} recovers Eq.~\eqref{Fetanormal} for $T\ge T_{\rm c}$. 
The contribution of thermal fluctuations to the free energy below $T_{\rm c}$ is obtained as
\beq
\bal
F_{\rm fl}^{-}&=-\frac{1}{\beta}\ln{\int \mathcal{D}[\eta({\bf r})] \, 
e^{-\beta F_{\eta}^{\rm Popov}[\eta({\bf r})]}} \\
&=-\frac{1}{2\beta}\sum_{\bf k}\Bigg[\ln{\left(\dfrac{\pi k_{\rm B}T}{a(T)+2b\langle\abs{\eta}^{2}\rangle+3b(\psi_{0}^{(\mathrm{GF})})^{2}+\gamma k^{2}}\right)} \\
&+\ln{\left(\dfrac{\pi k_{\rm B}T}{a(T)+2b\langle\abs{\eta}^{2}\rangle+b(\psi_{0}^{(\mathrm{GF})})^{2}+\gamma k^{2}}\right)}\Bigg].
\eal
\label{Fflbelow}
\eeq
The thermal average of the GF is self-consistently determined by
\beq
\bal
&\langle|{\eta}|^{2}\rangle=\dfrac{1}{\mathcal{Z}_{\eta}}\int \mathcal{D}[\eta({\bf r})]
|{\eta({\bf r})}|^{2}e^{-\beta F_{\eta}^{\rm Popov}[\eta({\bf r})]} \\
&=\dfrac{1}{2\beta L^{2}}\sum_{\bf k}\Bigg[\dfrac{1}{a(T)
+2b\langle|{\eta}|^{2}\rangle+3b(\psi_{0}^{({\rm GF})})^{2}+\gamma k^{2}}\\
&+\dfrac{1}{a(T)+2b\langle\abs{\eta}^{2}\rangle+b(\psi_{0}^{({\rm GF})})^{2}
+\gamma k^{2}}\Bigg] .
\eal
\label{eta2sf}
\eeq
Above $T_{\rm c}$ in which $\psi_{0}^{({\rm GF})}=0$, the two contributions 
from amplitude and phase fluctuations become identical and Eq.~\eqref{eta2sf} 
recovers Eq.~\eqref{eta2}. 
The temperature dependencies of $\langle\abs{\eta}^{2}\rangle$ and the temperature derivative are shown in Fig.~\ref{Figeta2}. 
We numerically find that $\langle\abs{\eta}^{2}\rangle$ including the GF determined by Eq.~\eqref{eta2} can exhibit a slight discontinuity at $T_{\rm c}$ as $\langle\abs{\eta}^{2}\rangle_{T\to T_{\rm c}^{-}}<\langle\abs{\eta}^{2}\rangle_{T\to T_{\rm c}^{+}}$. 
Nonetheless, the deviation $\langle\abs{\eta}^{2}\rangle_{T\to T_{\rm c}^{+}}-\langle\abs{\eta}^{2}\rangle_{T\to T_{\rm c}^{-}}$ is decreased with a smaller $\mathrm{Gi}$ and we set $\mathrm{Gi}=0.1\ll1$ so that the deviation is negligible. 
See Appendix \ref{Appeta} for details. 
The quasiorder parameter \eqref{phi0} including GF determined by Eq.~\eqref{eta2sf} is shown in Fig.~\ref{FigTc0TcTBKT} by a dashed green curve. 
With Eq.~\eqref{Tc0Tcdiff}, the quasiorder parameter \eqref{phi0} including the GF in the vicinity of $T_{\rm c}$ can be written as
\beq
(\psi_{0}^{(\mathrm{GF})})^{2}=\frac{\alpha k_{\rm B}}{b}A_{-}(T_{\rm c}-T),
\label{psiGFnearTc}
\eeq
with 
\beq
A_{\pm}\equiv 1+\dfrac{2b}{\alpha k_{\rm B}}\eval{\partial_{T}\langle\abs{\eta}^{2}\rangle}_{T\to T_{\rm c}^{\pm}}, 
\label{Aplusminus}
\eeq
which are directly related to the temperature derivative of $\langle\abs{\eta}^{2}\rangle$ shown in the inset of Fig.~\ref{Figeta2}. 
The coefficient $A_{-}$ in Eq.~\eqref{psiGFnearTc} reflects the Gaussian correction by $\langle\abs{\eta}^{2}\rangle$.

\section{Topological defects in the phase of the quasi-order parameter with Gaussian fluctuations}\label{SecBKTfl}

Let us write the space-dependent quasiorder parameter as 
\bseq
\beq
\bal
\psi({\bf r})&=\left[\psi_{0}^{(\text{BKT})}+\sigma({\bf r})\right]e^{i\theta_{0}({\bf r})+i\theta({\bf r})} \\
&\simeq \left[\psi_{0}^{(\text{BKT})}+\eta({\bf r})\right]e^{i\theta({\bf r})}, 
\label{psietavortex}
\eal
\eeq
\beq
\eta({\bf r})=\abs{\eta({\bf r})}e^{i\phi({\bf r})} ,
\eeq
\eseq
with $\eta({\bf r})=\sigma({\bf r})+i\theta_{0}({\bf r})\psi_{0}^{(\text{BKT})}$ including both amplitude and phase fluctuations to the first order where $\theta_{0}({\bf r})$ and $\phi({\bf r})$ being the regular phases with zero winding number. 
The phase field $\theta({\bf r})$ takes into account phase fluctuations with quantized vortices. 
The compactness of the phase angle field $\theta({\bf r})$ 
implies 
\beq 
\oint_\mathcal{C} {\boldsymbol \nabla} \theta({\bf r}) \cdot 
\mathrm{d}{\bf r} = 2\pi q 
\eeq
for any closed contour $\mathcal{C}$. 
Here, $q=0,\pm 1,\pm 2,\cdots$ is the integer topological charge associated with the corresponding quantized vortex (positive $q$) or antivortex (negative $q$). 
Inserting Eq.~\eqref{psietavortex} into Eq.~\eqref{gl}, one obtains
\beq
F_{\rm s}=F_{\mathrm{s}0}^{(\text{BKT})}+F_{\eta}+F_{\theta},
\eeq
with
\beq
\bal
&F_{\eta}\equiv \int\dd[2]{\bf r}\Bigg[\left[a(T)+2b(\psi_{0}^{(\text{BKT})})^{2}+\gamma(\grad\Tilde{\theta})^{2}\right]\abs{\eta({\bf r})}^{2}\\
&+\frac{b}{2}\abs{\eta({\bf r})}^{4} 
+\frac{b}{2}(\psi_{0}^{(\text{BKT})})^{2}\left[\eta({\bf r})^{2}+\eta^{*}({\bf r})^{2}\right] \\
&+b\psi_{0}\left[(\psi_{0}^{(\text{BKT})})^{2}+\abs{\eta({\bf r})}^{2}\right]\left[\eta({\bf r})+\eta^{*}({\bf r})\right] 
+\gamma\abs{\grad\eta({\bf r})}^{2}\Bigg],
\eal
\eeq
which recovers $F_{\eta}^{\rm Popov}$ in Eq.~\eqref{Fetapopov} under the Popov approximation by taking the thermal average and neglecting $(\grad\Tilde{\theta})^{2}$, and
\beq
\bal
&F_{\theta}\equiv
\gamma\int\dd[2]{\bf r}\Bigg[(\psi_{0}^{(\text{BKT})})^{2}(\grad\Tilde{\theta})^{2} \\
&+2\psi_{0}^{(\text{BKT})}\left(\grad\theta\right)^{2}\mathrm{Re}[\eta({\bf r})] 
+2\psi_{0}^{(\text{BKT})}\grad\theta\cdot\grad\mathrm{Im}[\eta({\bf r})] \\
&+\dfrac{1}{4(\psi_{0}^{(\text{BKT})})^{2}}\left[\eta({\bf r})\grad\eta^{*}({\bf r})-\eta^{*}({\bf r})\grad\eta({\bf r})\right]^{2}\Bigg], 
\eal
\label{FXY}
\eeq
which involves $\theta({\bf r})$ and 
\beq
\Tilde{\theta}({\bf r})\equiv\theta({\bf r})+\frac{\langle\abs{\eta}^{2}\rangle}{(\psi_{0}^{(\text{BKT})})^{2}}\phi({\bf r}). 
\label{thetatilde}
\eeq 
The last line in Eq.~\eqref{FXY} is the higher-order contribution in $\eta({\bf r})$ which is neglected in the following. 

From Eq.~\eqref{FXY}, the equation of motion in terms of $\theta$ reads
\beq
\bal
0&=-\gamma(\psi_{0}^{(\text{BKT})})^{2}(\grad\Tilde{\theta})^{2}+\gamma\psi_{0}^{(\text{BKT})}\left[\eta({\bf r})+\eta^{*}({\bf r})\right]\grad^{2}\theta \\
&-i\gamma\psi_{0}^{(\text{BKT})}\left[\grad^{2}\eta^{*}({\bf r})-\grad^{2}\eta({\bf r})\right] . 
\eal
\label{EoMtheta}
\eeq
By performing the thermal average with respect to $\eta$ on the equation of motion \eqref{EoMtheta}, one may obtain
\beq
\bal
0=\gamma(\psi_{0}^{(\text{BKT})})^{2}\grad^{2}\Tilde{\theta} 
=\fdv{F_{\theta}^{\rm Popov}}{\theta},
\eal
\label{EoMthetaav}
\eeq
where
\beq
F_{\theta}^{\rm Popov}\equiv \frac{J_{0}(T)}{2}\int\dd[2]{\bf r}(\grad\Tilde{\theta})^{2}
\label{FXYpopov}
\eeq
is the XY free energy under the Popov approximation generating Eq.~\eqref{EoMthetaav} with $J_{0}(T)\equiv 2\gamma(\psi_{0}^{(\text{BKT})})^{2}$ being the bare phase stiffness. 
Note that $\Tilde{\theta}({\bf r})$ in Eq.~\eqref{thetatilde} keeps the circulation of vorticity invariant from that of $\theta({\bf r})$ because $\phi({\bf r})$ is regular. 
The equation of motion with respect to $\psi_{0}^{(\text{BKT})}$ reads
\beq
\bal
0&=\fdv{F_{\rm s}}{\psi_{0}^{(\text{BKT})}} \\
&=2\psi_{0}^{(\text{BKT})}\left[a(T)+b(\psi_{0}^{(\text{BKT})})^{2}+2b\abs{\eta({\bf r})}^{2}+\gamma(\grad\Tilde{\theta}({\bf r}))^{2}\right]. 
\eal
\label{eompsi0BKT}
\eeq
Here, we neglected the anomalous terms of $\eta({\bf r})$ which vanish with the thermal average. 
Performing the thermal average with respect to $F_{\rm s}^{\rm Popov}=F_{\mathrm{s}0}^{(\text{BKT})}+F_{\eta}^{\rm Popov}+F_{\theta}^{\rm Popov}$ with
\beq
\bal
\frac{F_{\mathrm{s}0}^{(\text{BKT})}}{L^{2}}&=\left[a(T)+2b\langle\abs{\eta}^{2}\rangle+\gamma\langle(\grad\Tilde{\theta})^{2}\rangle\right](\psi_{0}^{(\text{BKT})})^{2} \\
&+\frac{b}{2}(\psi_{0}^{(\text{BKT})})^{4},
\eal
\label{uniform-bkt}
\eeq
one obtains
\beq
(\psi_{0}^{(\text{BKT})})^{2}=-\dfrac{a(T)+2b\langle\abs{\eta}^{2}\rangle+\gamma\langle(\grad\Tilde{\theta})^{2}\rangle}{b}, 
\label{psi0BKT}
\eeq
if $a(T)+2b\langle\abs{\eta}^{2}\rangle+\gamma\langle(\grad\Tilde{\theta})^{2}\rangle\ge0$
where $\langle\abs{\eta}^{2}\rangle$ is given by $\langle\abs{\eta}^{2}\rangle=\int\mathcal{D}[\eta({\bf r})]\abs{\eta({\bf r})}^{2}\mathrm{exp}[-\beta F_{\eta}^{\rm Popov}[\eta({\bf r})]]/\mathcal{Z}_{\eta}$ and $\langle(\grad\Tilde{\theta})^{2}\rangle$ can be evaluated as
\beq
\bal
\langle(\grad\Tilde{\theta})^{2}\rangle&=\frac{1}{\mathcal{Z}_{\theta}}\int\mathcal{D}[\Tilde{\theta}({\bf r})](\grad\Tilde{\theta}({\bf r}))^{2}\,e^{-\beta F_{\theta}^{\rm Popov}} \\
&=\frac{\Lambda^{2}-k_{0}^{2}}{4\pi K_{0}(T)},
\eal
\label{theta2}
\eeq
with $\mathcal{Z}_{\theta}\equiv \int\mathcal{D}[\Tilde{\theta}({\bf r})]e^{-\beta F_{\theta}^{\rm Popov}}$ and $K_{0}(T)=\beta J_{0}(T)$. 
In the high-temperature regime in which $a(T)+2b\langle\abs{\eta}^{2}\rangle+\gamma\langle(\grad\Tilde{\theta})^{2}\rangle<0$, $\psi_{0}^{(\text{BKT})}=0$. 
Inserting Eq.~\eqref{theta2} into Eq.~\eqref{psi0BKT}, we determine $\psi_{0}$ in Eq.~\eqref{psi0BKT} and the bare superfluid phase stiffness $J_{0}(T)$. 
By solving Eq.~\eqref{theta2} with respect to $\langle(\grad\Tilde{\theta})^{2}\rangle$, one obtains $\langle(\grad\Tilde{\theta})^{2}\rangle=[-a(T)-2b\langle\abs{\eta}^{2}\rangle-\sqrt{D}]/2\gamma$.
Unfortunately, it has a real root only in the low-temperature regime in which $D\equiv[-a(T)-2b\langle\abs{\eta}^{2}\rangle]^{2}-b(\Lambda^{2}-k_{0}^{2})/2\pi\beta\ge0$ holds. 
Then, in the high-temperature regime in which $D<0$, we assume $\langle(\grad\Tilde{\theta})^{2}\rangle=[-a(T)-2b\langle\abs{\eta}^{2}\rangle]/2\gamma$, which vanishes at $T_{\rm c}$ by neglecting the imaginary part. 
Consequently, in the high-temperature regime of $D<0$, we have $\psi_{0}^{(\text{BKT})}=\psi_{0}^{(\mathrm{GF})}/\sqrt{2}$. 

As shown by Kosterlitz and Thouless \cite{kosterlitz}, in the two-dimensional case, the total number of quantized vortices varies as a function of the temperature: at zero temperature there are no vortices, however as the temperature increases vortices start to appear in vortex-antivortex pairs. 
The phase stiffness is renormalized by the presence of these vortex-antivortex pairs. The pairs are bound at low temperatures, until at the BKT critical temperature $T_{\rm BKT}$ an unbinding transition occurs above which a proliferation of free vortices and antivortices is observed. At $T_{\rm BKT}$ the renormalized phase stiffness $J_{\rm R}(T)$ jumps to zero and for $T>T_{\rm BKT}$ the superfluidity is completely lost. 
Moreover, the renormalized free energy becomes
\beq
\bal 
F_{\rm s}^{\rm Popov} &= F_{\mathrm{s}0}^{(\text{BKT})}[\psi_{\rm R}^{\rm BKT}]+F_{\eta}^{\text{Popov}}[\psi_{\rm R}^{\rm BKT}] \\
&+\int\dd[2]{\bf r}{J_{\rm R}(T)\over 2}|\grad\Tilde{\theta}({\bf r})|^2 ,
\eal
\label{free-with-bkt}
\eeq
where 
\beq 
\psi_{\rm R}^{({\rm BKT})}(T)=\sqrt{J_{\rm R}(T)\over 2\gamma} .
\eeq

We now discuss how one can obtain the renormalized phase stiffness $J_{\rm R}(T)$ from the bare one $J_0(T)$. 
The NK renormalization group equations are given by \cite{kosterlitz}
\beqa 
\partial_{l}K_l(T)^{-1} &=& 4\pi^3 y_l(T)^2  ,
\label{rg1}
\\
\partial_{l} y_l(T) &=& \left[2 - \pi K_l(T)\right]y_l(T) ,
\label{rg2}
\eeqa
where $l$ is the running RG scale, which goes from $l=0$ (bare results) to $l=l_{\rm max}=\ln{(\pi/\xi k_{0})}$ (fully renormalized results), 
\beq
J_l(T) = k_{\rm B} T \, K_l(T) 
\eeq
is the running phase stiffness, and 
\beq 
\mu_l(T) = - k_{\rm B} T \, \ln{y_l(T)}
\eeq
is the running vortex-core energy. 
We can obtain the fully renormalized phase stiffness $J_{\rm R}(T)=J_{l_{\rm max}}(T)$ and the fully renormalized vortex-core energy $\mu_{\rm R}(T)=\mu_{l_{\rm max}}(T)$. 

Quite remarkably, by separation of variables, from Eqs.~\eqref{rg1} and \eqref{rg2} one finds \cite{stoof} 
\beq 
y_l(T)^2 - {1\over \pi^3 K_l(T)} - {1\over 2\pi^2} 
\ln{K_l(T)} = C \; , 
\label{henk}
\eeq
where $C$ is an integration constant, determined by the initial conditions. 
For the critical trajectory, where $T=T_{\rm BKT}$, one has $y_{\rm R}(T_{\rm BKT})=0$ and $K_{\rm R}(T_{\rm BKT})=2/\pi$, namely, 
\beq 
k_{\rm B}T_{\rm BKT} = {\pi\over 2} J_{\rm R}(T_{\rm BKT}^{-}) \; . 
\label{nk-exact}
\eeq
In addition, from Eq.~\eqref{henk} and the previous expressions, 
one also obtains for the critical trajectory \cite{stoof}
\beq 
C={\ln(\pi/2)-1\over 2\pi^2} = -0.0278 \; . 
\label{C-one}
\eeq

It is important to stress that Eq.~\eqref{nk-exact} relates the BKT critical temperature $T_{\rm BKT}$ 
to the fully renormalized phase stiffness $J_{\rm R}(T_{\rm BKT})$ calculated at the same BKT critical temperature. 
Sometimes, one uses an approximated version of Eq.~\eqref{nk-exact}, the so-called NK criterion \cite{nelson}, substituting $J_{\rm R}(T_{\rm BKT}^{-})$ with $J_0(T_{\rm BKT}^{-})$. 
In this way, one quickly gets an approximated $T_{\rm BKT}$ from the knowledge of the bare superfluid density $J_0(T)$. 
Here, however, we improve this approximation. 
For the initial condition of the vortex core energy we choose 
\beq 
\mu_{0}(T) = {\pi^2\over 4} J_0(T) \; , 
\eeq
that is currently the most rigorous choice for superconductors and 
superfluids \cite{pip1,pip2,pip3,pip4}. 
As a consequence, the initial condition of the vortex fugacity reads 
\beq 
y_0(T)= e^{-\pi^2 K_0(T)/4} \; . 
\label{giov}
\eeq
Consequently, at the BKT critical temperature from Eqs.~\eqref{henk}, 
\eqref{C-one}, and \eqref{giov} we have 
\beq
\bal
&{2\over \pi K_0(T_{\rm BKT})}-\ln{\left(\dfrac{2}{\pi K_0(T_{\rm BKT})}\right)}-2\pi^{2}e^{-\pi^2 K_0(T_{\rm BKT})/2} \\
&=1  \; .  
\eal
\eeq
The unknown $K_0(T_{\rm BKT})$ can be then found by solving 
numerically the above equation, obtaining 
\beq 
K_0(T_{\rm BKT}) = 1.055 \;  
\eeq
or, equivalently, from $K_0(T_{\rm BKT})=J_0(T_{\rm BKT})/(k_{\rm B}T_{\rm BKT})$, 
the remarkable semi-analytical result \cite{midei2024a}
\beq 
k_{\rm B}T_{\rm BKT} = 0.948 \, J_0(T_{\rm BKT}) \; . 
\eeq
This formula is much simpler than Eq.~\eqref{nk-exact} because it depends only on the knowledge of the bare phase stiffness $J_0(T)$. 
Moreover, it is much more reliable than the approximated NK criterion discussed earlier. 

Several theoretical studies beyond perturbative RG analysis revealed that the RG equations \eqref{rg1} and \eqref{rg2} are not subject to modification due to the coupling with the Higgs mode by redefining the vortex fugacity including a constant factor \cite{wetterich1995,wetterich2001,delamotte2014,metzner2017,kopietz2017,defenu2017}. 
Based on this knowledge, we keep using the RG equations but with the bare quantities given by $K_{0}(T)=2\gamma(\psi_{0}^{(\text{BKT})})^{2}/k_{\rm B}T$ and $y_{0}(T)=\mathrm{exp}[-\pi^{2}K_{0}(T)/4]$ determined from Eq.~\eqref{psi0BKT} which encodes the coupling with the amplitude mode. 

Using the renormalized phase stiffness $K_{\rm R}(T)=2\gamma (\psi_{\rm R}^{(\text{BKT})})^{2}/k_{\rm B}T$ and vortex fugacity $y(T)$, we can obtain the effective free energy by integrating out $\Tilde{\theta}$ in Eq.~\eqref{FXYpopov} as
\beq
\bal
F_{\rm fl,XY}&=-\frac{1}{\beta}\ln{\int\mathcal{D}[\Tilde{\theta}({\bf r})] \, e^{-\beta F_{\theta}^{\rm Popov}}} \\
&=-\frac{1}{\beta}\sum_{\bf k}\ln{\left(\dfrac{2\pi}{K_{\rm R}(T)k^{2}}\right)}.
\eal
\label{FflXY}
\eeq
As in Eq.~\eqref{Fflbelow}, the integration of $\eta(\bf{r})$ results in
\begin{widetext}
\beq
\bal
F_{\rm fl}^{-}&=-\frac{1}{\beta}\ln{\int \mathcal{D}[\eta({\bf r})] \, 
e^{-\beta F_{\eta}^{\rm Popov}[\eta({\bf r})]}} \\
&=-\frac{1}{2\beta}\sum_{\bf k}\Bigg[\ln{\left(\dfrac{\pi k_{\rm B}T}{a(T)+2b\langle\abs{\eta}^{2}\rangle+\gamma\langle(\grad\Tilde{\theta})^{2}\rangle+3b(\psi_{\rm R}^{(\text{BKT})})^{2}+\gamma k^{2}}\right)} \\
&+\ln{\left(\dfrac{\pi k_{\rm B}T}{a(T)+2b\langle\abs{\eta}^{2}\rangle+\gamma\langle(\grad\Tilde{\theta})^{2}\rangle+b(\psi_{\rm R}^{(\text{BKT})})^{2}+\gamma k^{2}}\right)}\Bigg].
\eal
\label{FfletaBKT}
\eeq
\end{widetext}
In the normal phase in which $\psi_{\rm R}^{(\text{BKT})}=0$, $F_{\rm fl}^{+}$ is given by Eq.~\eqref{Fflabove} and we neglect $F_{\rm fl, XY}$ because we have $F_{\theta}[\psi_{0}=\psi_{\rm R}^{(\text{BKT})}]=0$. 

\begin{figure}[t]
\centering
\includegraphics[keepaspectratio,scale=0.35]{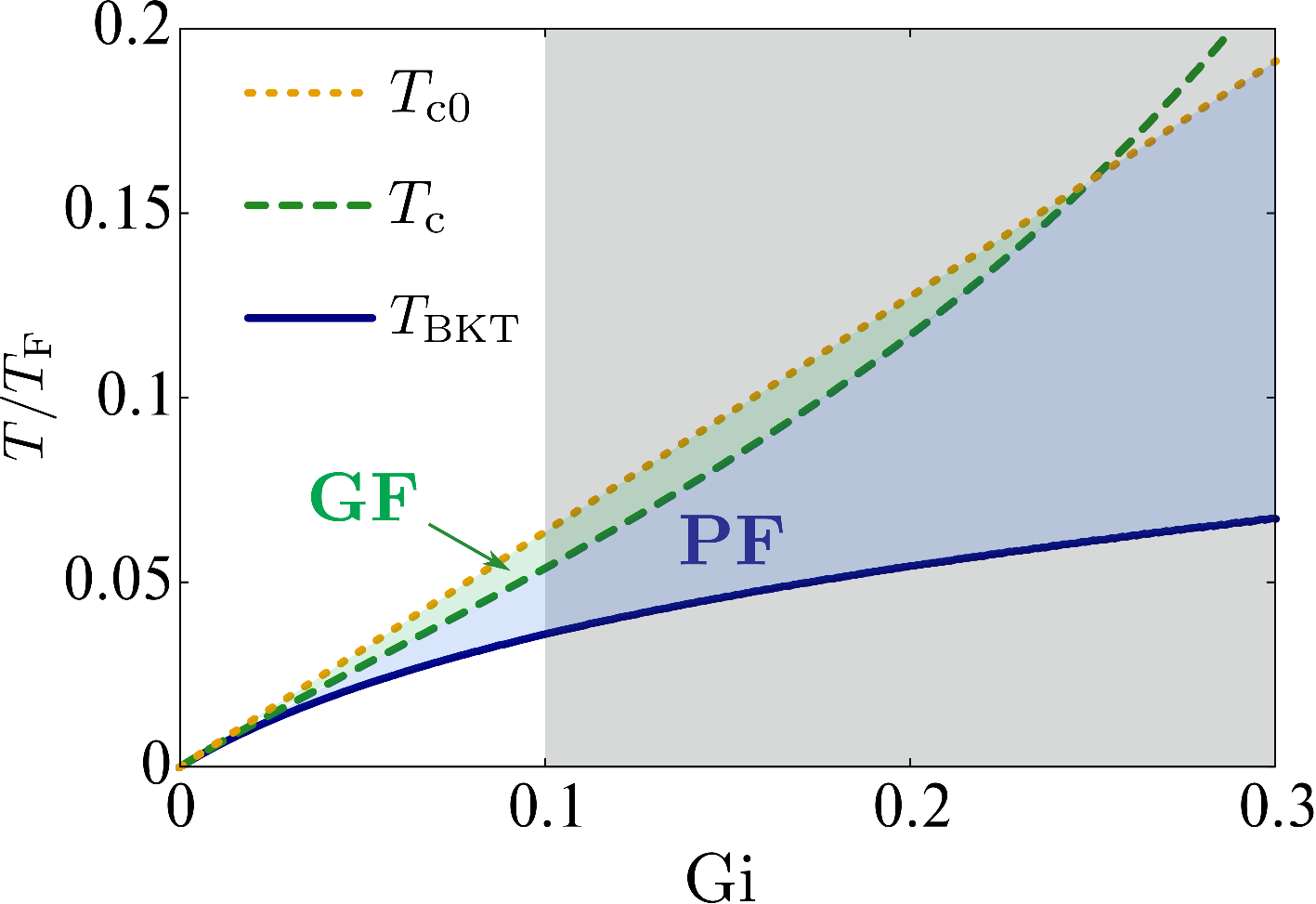}
\caption{Dependence of the critical temperatures $T_{\mathrm{c}0}$, $T_{\rm c}$, and $T_{\rm BKT}$ in the unit of $T_{\rm F}$ on $\mathrm{Gi}$. 
The gray region $\mathrm{Gi}\gtrsim 0.1$ is the unphysical region in which $T_{\rm c}/T_{\mathrm{c}0}$ increases in terms of $\mathrm{Gi}$. 
The shaded region of $T_{\rm c}<T<T_{\mathrm{c}0}$ is governed by the Gaussian thermal fluctuations (GF), while the phase fluctuations (PF) associated with vortex excitations are dominant in the region of $T_{\rm BKT}<T<T_{\mathrm{c}}$. }
\label{FigTBKT}
\end{figure}

The superfluid phase transition temperatures scaled by the Fermi temperature are summarized in Fig.~\ref{FigTBKT}. 
In Fermi systems, a larger value of $\mathrm{Gi}=\pi T_{\mathrm{c}0}/2T_{\rm F}$ corresponds to a stronger pairing interaction. 
In our Ginzburg-Landau problem, when the bare phase stiffness in the vicinity of $T_{\rm c}$ is given by
\beq
J_{0}(T)=\dfrac{k_{\rm B}}{2\pi \mathrm{Gi}}A_{-}(T_{\rm c}-T), 
\label{bareJA}
\eeq
we obtain 
\beq 
\bal
{T_{\rm c} - T_{\rm BKT} \over T_{\rm BKT} } &= \dfrac{2\pi}{0.948 A_{-}} \, \mathrm{Gi}   
\eal
\label{shiftbkt}
\eeq
with $\mathrm{Gi}$ the Ginzburg-Levanyuk number defined by Eq.~\eqref{levanyuk} 
and $A_{-}$ is defined by Eq.~\eqref{Aplusminus}. 
Equation \eqref{shiftbkt} is valid if the BKT transition temperature is close to $T_{\rm c}$ at which the linear approximation of the phase stiffness in Eq.~\eqref{bareJA} holds, corresponding to $\mathrm{Gi}\ll1$.  
Without Gaussian fluctuations, Eq.~\eqref{shiftbkt} reduces to 
\beq
\frac{T_{\mathrm{c}0}-T_{\mathrm{BKT},0}}{T_{\mathrm{BKT},0}}=\frac{2\pi}{0.948}\mathrm{Gi}, 
\eeq
where $T_{\mathrm{BKT},0}$ represents the BKT transition temperature without the Gaussian thermal fluctuations. 
The dotted-dashed gray curve in Fig.~\ref{FigTc0TcTBKT} represents the quasiorder parameter without Gaussian fluctuations. 
The obtained BKT transition temperature \eqref{shiftbkt} is lower than the upper bound $T_{\rm F}/8$ for $\mathrm{Gi}\lesssim0.4$ under $A_{-}\simeq1$ \cite{shi2023}. 
Figure \ref{FigTBKT} also reveals that the ratio between the GF and MF critical temperatures $T_{\rm c}/T_{\mathrm{c}0}$ swells with a large $\mathrm{Gi}$, which is unphysical because fluctuations must tend to break ordered phase lowering the transition temperature. 
The gray region in Fig.~\ref{FigTBKT} represents the unphysical region in which $T_{\rm c}/T_{\mathrm{c}0}$ increases with respect to $\mathrm{Gi}$. 
Therefore, we can identify $\mathrm{Gi}\lesssim0.1$ as the regime in which our Ginzburg-Landau analysis is physically reasonable. 

The effects of fluctuations on the transition temperature are summarized in Fig.~\ref{FigTc0TcTBKT}. 
It shows the quasiorder parameters $|\psi_{0}|^{2}=\alpha k_{\rm B}(T_{\mathrm{c}0}-T)/b$ within the MF level in Eq.~\eqref{psi0}, $|\psi_{0}^{\mathrm{({\rm GF})}}|^{2}=-a(T)/b-2\langle|\eta|^{2}\rangle$ at the GF level, and $|\psi_{\rm R}^{({\rm BKT})}|^{2}=J_{\rm R}(T)/2\gamma$ including the vortex excitations. 
Including the GF without vortex excitations decreases the $T_{\mathrm{c}0}$ to $T_{\rm c}$ as in Eq.~\eqref{shift2d}. 
The PF associated with the vortex excitations further reduce the critical temperature from $T_{\rm c}$ to $T_{\mathrm{BKT},0}$, and the coupling with the amplitude mode reduces $T_{\mathrm{BKT},0}$ to $T_{\rm BKT}$ as reported in Fig.~\ref{FigTc0TcTBKT}.

\section{Magnetic properties}\label{SecHc}

In this section, we show the magnetic properties by focusing on the penetration depth and the critical magnetic fields. 

\subsection{Penetration depth}

Adopting the saddle-point approximation with respect to the appropriate free energy functional, one can deduce the London penetration depth \cite{sala2020,annett}
\beq
\lambda(T)=\sqrt{\dfrac{b}{2\mu_{0}Q^{2}\gamma a(T)}}
=\sqrt{\dfrac{1}{2\mu_{0}Q^{2}\gamma \abs{\psi_{0}(T)}^{2}}} 
\label{lambda}
\eeq
of the magnetic field at the MF level, at the GF level, 
and also at the BKT level by substituting $\psi_{0}$ with $\psi_{0}^{({\rm GF})}$ and $\psi_{\rm R}^{({\rm BKT})}$ where $Q$ is the electrical charge and $\mu_{0}$ is the magnetic constant. 

Having the London penetration depth $\lambda(T)$ and the Ginzburg-Landau 
coherence length 
\beq 
\xi(T) = \sqrt{\gamma\over |a(T)|} 
=\sqrt{\dfrac{\gamma}{b\abs{\psi_{0}(T)}^{2}}},
\label{xibelow}
\eeq
one immediately finds the Ginzburg-Landau parameter 
\beq 
\kappa = {\lambda(T) \over \xi(T)} =\sqrt{\dfrac{b}{2\mu_{0}Q^{2}\gamma^{2}}}\; , 
\label{kappa}
\eeq
such that $\kappa < 1/\sqrt{2}$ for type-I 
superconductors and $\kappa > 1/\sqrt{2}$ for type-II superconductors 
\cite{annett}.  
We determine the quasiorder parameters in Eqs.~\eqref{lambda} and \eqref{xibelow} by the saddle-point of the free energy in Eq.~\eqref{uniform} within the MF level, Eq.~\eqref{uniform-gf} including GF, and Eqs.~\eqref{rg1} and \eqref{rg2} including the vortex excitations responsible for the BKT transition. 
The GL parameter is independent of the order parameter and temperature by definition of Eq.~\eqref{kappa}. 
BKT transitions can be discussed in type-II superconductors and we focus on them in the following \cite{timm}. 

The coherence length in Eq.~\eqref{xibelow} also represents the vortex core size. 
According to Fig.~\ref{FigTc0TcTBKT}, Eq.~\eqref{xibelow} implies that the vortex core size becomes larger by taking into account thermal fluctuations. 

\subsection{Critical magnetic field}

\begin{figure}[t]
\centering
\includegraphics[keepaspectratio,scale=0.35]{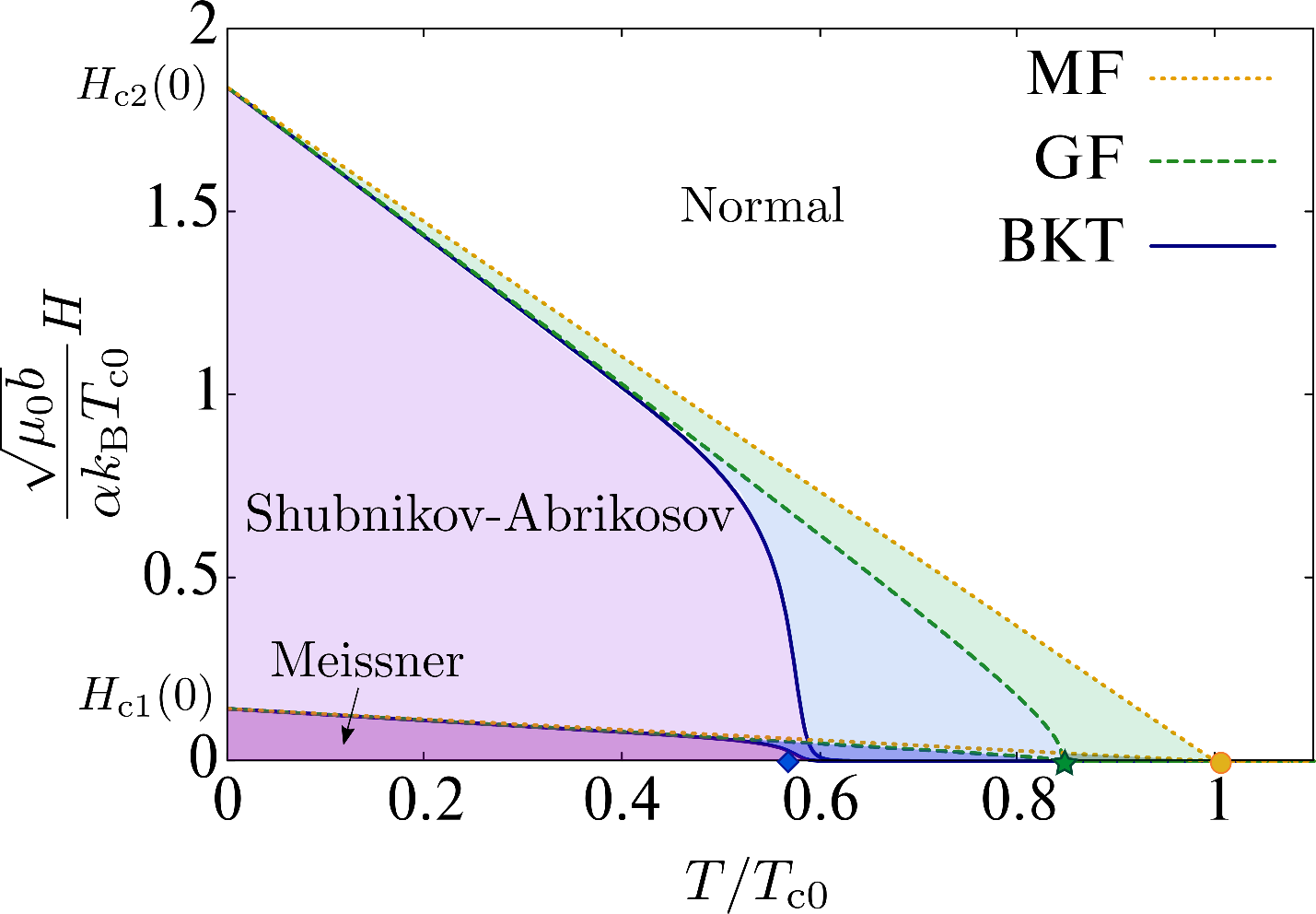}
\caption{$H$-$T$ phase diagram of a type-II superconductor within the MF level, including the GF, and the vortex excitations, respectively. 
The critical magnetic fields are given by Eqs.~\eqref{Hc1Hc2} with Eqs.~\eqref{HcMF}, \eqref{Hc0}, and \eqref{Hc0BKT}, respectively, for 
$\mathrm{Gi}=0.1$ and $\kappa=1.3$. 
The phase in the region $0<H<H_{\mathrm{c}1}$ is the Meissner phase while the one in $H_{\mathrm{c}1}<H<H_{\mathrm{c}2}$ is the Shubnikov-Abrikosov phase \cite{annett}. }
\label{FigHc1Hc2}
\end{figure}

At the MF level, the thermodynamic critical magnetic field $H_{\rm c}$ is related to the free energy as \cite{annett}
\beq
F_{\rm s0}=-L^{2}\dfrac{\mu_{0}}{2}H_{\rm c}^{2}.  
\label{FsHc}
\eeq
It follows that the critical magnetic field reads \cite{annett}
\beq
\bal
H_{\rm c}(T)&=\sqrt{\dfrac{2}{\mu_{0}}
\abs{a(T)\psi_{0}^{2}(T)+\dfrac{b}{2}\psi_{0}(T)^{4}}} \\
&=\sqrt{\dfrac{b}{\mu_{0}}}\psi_{0}(T)^{2}
=\dfrac{\alpha k_{\rm B}}{\sqrt{\mu_{0}b}}(T_{\mathrm{c}0}-T). 
\eal
\label{HcMF}
\eeq

We now include the effects of fluctuations in the calculation of the critical magnetic field. 
However, adopting the saddle-point approximation, we use instead Eq.~\eqref{uniform-gf} as
\beq 
F_{\rm s0}^{({\rm GF})}=-L^{2}\dfrac{\mu_{0}}{2} 
(H_{\rm c}^{({\rm GF})})^{2}, 
\eeq
and the critical magnetic field is given by
\beq
\bal
H_{\rm c}^{({\rm GF})}(T)  
=\sqrt{\dfrac{b}{\mu_{0}}}\left(\psi_{0}^{({\rm GF})}\right)^{2}\; ,
\eal
\label{Hc0}
\eeq
with Eq.~\eqref{phi0}. 
By including the vortex excitations and adopting a similar saddle-point 
approximation on the BKT free energy \eqref{uniform-bkt}, the critical magnetic field 
is subject to renormalization as
\beq
\bal
H_{\rm c}^{({\rm BKT})}(T)
=\sqrt{\dfrac{b}{\mu_{0}}}\left(\psi_{\rm R}^{({\rm BKT})}\right)^{2}\; .
\eal
\label{Hc0BKT}
\eeq

With the Ginzburg-Landau parameter $\kappa$, the critical magnetic fields $H_{\mathrm{c}1}$ and $H_{\mathrm{c}2}$ of type-II superconductors are deduced from $H_{\rm c}$ by using the formulas \cite{annett}
\bseq
\beq
H_{\mathrm{c}1}(T)= {1\over \sqrt{2}} {\ln(\kappa)\over \kappa} \, H_{\rm c}(T), 
\label{Hc1}
\eeq
\beq
H_{\mathrm{c}2}(T) = \sqrt{2} \kappa \, H_{\rm c}(T) \; . 
\label{Hc2}
\eeq
\label{Hc1Hc2}
\eseq
Here, the expression of Eq.~\eqref{Hc1} is valid only with $\kappa\gg1/\sqrt{2}$ \cite{annett}. 
Figure \ref{FigHc1Hc2} shows the $H$-$T$ phase diagram within the MF level and the beyond-MF level including GF, and vortex excitations, respectively, for $\mathrm{Gi}=0.1$. 
We used the GL parameter $\kappa=1.3$, which is the typical value in $\rm Nb$ superconductors \cite{annett}. 
The phase in the region $0<H<H_{\mathrm{c}1}$ is the Meissner phase while the one in $H_{\mathrm{c}1}<H<H_{\mathrm{c}2}$ is the Shubnikov-Abrikosov phase \cite{annett}.

\section{Heat capacity}\label{Secheat}

From the free energy, one can also determine the size of the discontinuous jump in the heat capacity at the superfluid phase transition point as \cite{larkin}
\beq
\Delta C=-\dfrac{T}{L^{2}}\pdv[2]{F_{\rm s}}{T}.
\label{DeltaC}
\eeq
First, we start with a recap of the mean-field case plus the thermal fluctuations, which is exactly the case considered in Ref.~\cite{larkin}. 
The contribution from the thermal fluctuations $F_{\rm fl}$ can be singular while that from $F_{\mathrm{s}0}$ is regular. 
In the vicinity of $T_{\mathrm{c}0}$, the heat capacity due to the thermal fluctuations reads
\beq
\bal
\Delta C_{\rm fl,MF}^{\pm}=-\dfrac{T}{L^{2}}\pdv[2]{F_{\rm fl,MF}^{\pm}}{T} .
\eal
\label{CMFpm}
\eeq
By focusing only on the most singular part in Eq.~\eqref{CMFpm} in the vicinity of the superfluid phase transition temperature, a straightforward calculation yields (see Appendix \ref{AppCv})
\beq
\Delta C_{\rm fl,MF}^{-}=\Delta C_{\rm fl,MF}^{+}, 
\label{DeltaCflMFrelation}
\eeq
if $\xi\Lambda\to\infty$ and $\xi k_{0}\to0$. 
Otherwise, the relation \eqref{DeltaCflMFrelation} no longer holds. 
Consequently, the heat capacity exhibits no jump at the superfluid phase transition under $\mathrm{Gi}\to0$ within the mean-field level in 2D \cite{larkin}. 

By including the GF, we obtain
\beq
\Delta C_{\rm fl,GF}^{-}
=\left(\dfrac{A_{-}}{A_{+}}\right)^{3}\Delta C_{\rm fl,GF}^{+}, 
\label{DeltaCflGFrelation}
\eeq
if $\xi\Lambda\to\infty$ and $\xi k_{0}\to0$. 
Otherwise, the relation \eqref{DeltaCflGFrelation} no longer holds. 
In contrast to the MF case in Eq.~\eqref{DeltaCflMFrelation}, Eq.~\eqref{DeltaCflGFrelation} implies that the heat capacity can exhibit discontinuity at the superfluid transition point even with $\xi\Lambda\to\infty$ and $\xi k_{0}\to0$ due to the Gaussian corrections in $A_{\pm}$.  

\begin{figure}[t]
\centering
\includegraphics[keepaspectratio,scale=0.3]{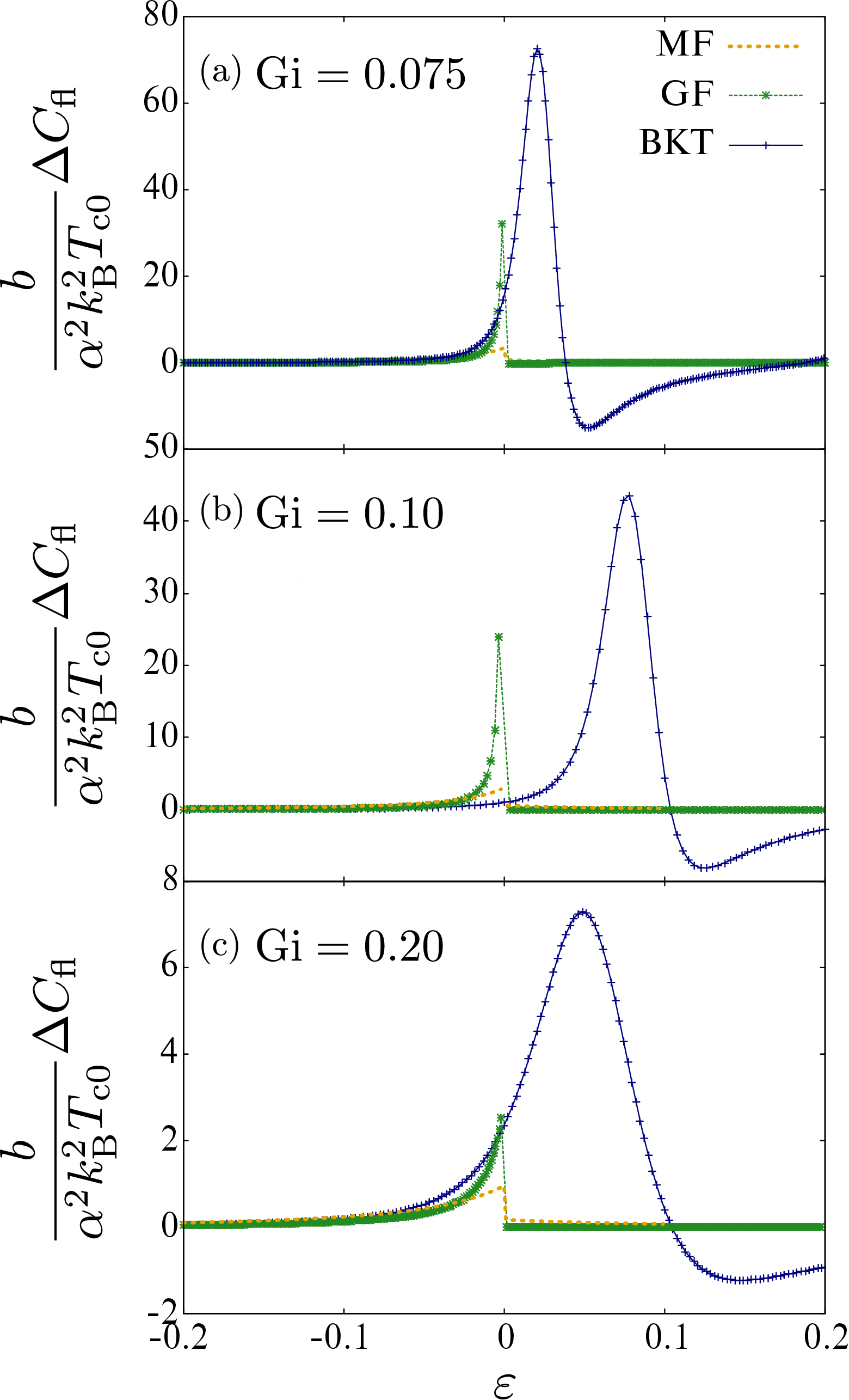}
\caption{Fluctuation contributions to the heat capacity $\Delta C_{\rm fl}$ as a function of the reduced temperature $\varepsilon=(T-T_{\mathrm{c}0})/T_{\mathrm{c}0} \,\text{(MF)},\,(T-T_{\mathrm{c}})/T_{\mathrm{c}} \,\text{(GF)}, (T-T_{\mathrm{BKT}})/T_{\mathrm{BKT}} \,\text{(BKT)}$, with $\mathrm{Gi}=0.075 \,(\mathrm{a}),\,0.1 \,(\mathrm{b}),\,0.2 \,(\mathrm{c})$. }
\label{FigCfl}
\end{figure}

Figure \ref{FigCfl} shows the heat capacity due to the fluctuation contribution $\Delta C_{\rm fl}=-T\partial^{2} F_{\rm fl}^{\pm}/\partial T^{2}/L^{2}$ with three different values of the Ginzburg-Levanyuk number $\mathrm{Gi}$, which may exhibit singular behavior at the transition temperature, with respect to the reduced temperature 
\beq
\varepsilon=
\begin{cases}
\displaystyle\frac{T-T_{\mathrm{c}0}}{T_{\mathrm{c}0}} & (\text{MF}),\\\\
\displaystyle\frac{T-T_{\mathrm{c}}}{T_{\mathrm{c}}} & (\text{GF}),\\\\
\displaystyle\frac{T-T_{\mathrm{BKT}}}{T_{\mathrm{BKT}}} & (\text{BKT}).
\end{cases}
\eeq
In both MF and GF levels, the heat capacity exhibits a discontinuity at criticality $\varepsilon=0$. 
This is consistent with our analytic calculation in Appendix \ref{AppCv}. 
Indeed, we are working with a finite momentum cutoff $k_{0}>0$ and the continuous behavior of the heat capacity in a 2D superconductor within the MF level can be obtained only under the condition $\xi k_{0}\to0$ as well as $\xi\Lambda\to\infty$. 
However, by taking into account the BKT transition, the heat capacity shows no singularity at $\varepsilon=0$. 
This continuous change originates from the smooth behavior of the quasiorder parameter at $T_{\rm BKT}$ shown in Fig.~\ref{FigTc0TcTBKT} and reflects the infinite-order phase transition \cite{altlandsimons}. 
Instead, it exhibits a maximum above the BKT transition temperature for any values of $\mathrm{Gi}$, which is consistent with a numerical prediction \cite{altlandsimons}. 
The contribution of phase fluctuations in Eq.~\eqref{FflXY} is mainly responsible for the behavior of $\Delta C_{\rm fl}$. 
In other words, the measurement of the heat capacity around the critical temperature can be useful to verify the BKT transition in a finite-size superconducting material \cite{nguyen2020}. 
For instance, Mizukami {\it et al.} reported a BCS-like jump of the heat capacity at the superconducting transition temperature in the nematic phase of $\mathrm{FeSe}_{1-x}\mathrm{S}_{x}$, when $x<0.17$ \cite{shibauchi2023}. 
This can be attributed to the dimensionality of the sample that is not strictly 2D and thus, Gaussian fluctuations are dominating over the topological ones giving the typical jump in the heat capacity at the transition temperature. 
We suggest that the change in behavior of the heat capacity can be used to detect the transition from the 3D to the 2D regime when the thickness of the sample is reduced. 
However, upon entering the tetragonal phase at $x>0.17$, where nematic order is suppressed, the discontinuous jump of the heat capacity at the critical temperature disappears. 
This has been attributed to a reminiscence of the BEC transition in Bose gas systems, so in this regime, the continuity of the heat capacity at the transition is due to other effects rather than dimensionality. 
In this phase, highly non-mean-field behaviors consistent with BEC-like pairing are found in the thermodynamic quantities with giant superconducting fluctuations extending far above $T_{\rm c}$. 
The broadening of the peak in the heat capacity at the superconducting transition when the system is tuned toward the crossover and the BEC regime of the BCS-BEC crossover observed in experiments is compatible with our results in Fig.~\ref{FigCfl}. 
A larger $\mathrm{Gi}$ decreases the RG cutoff $l_{\rm max}=\ln{(\pi/\xi k_{0})}$ smearing the jump of the phase stiffness, which results in the broadening of the heat capacity as shown in Fig.~\ref{FigCfl}.  
With $\mathrm{Gi}=0.075$ shown in Fig.~\ref{FigCfl}(a), the maximum of the BKT heat capacity is closer to $\varepsilon=0$ than the one with $\mathrm{Gi}=0.1$ plotted in Fig.~\ref{FigCfl}(b) and the behaviors of the GF and BKT heat capacities in the superfluid phase $\varepsilon<0$ are almost indistinguishable. 
However, the maximum of the BKT heat capacity above $T_{\rm BKT}$ makes the behaviors in the normal phase $\varepsilon>0$ dramatically different. 
Compared with Fig.~\ref{FigCfl}(c), we can also observe that the heat capacity is suppressed as well as broadened by a large value of $\mathrm{Gi}$. 
It demonstrates that a smaller infrared momentum cutoff $k_{0}=(\mathrm{Gi}\alpha k_{\rm B}T_{\rm c}/\gamma)^{1/2}$ enhances the heat capacity not only at the GF level [see Eqs.~\eqref{DeltaCsfn2Dfull}] but also at the BKT level.

\section{Conclusions}\label{Secconclusion}

In conclusion, we performed the Ginzburg-Landau analysis to reveal that the mean-field critical temperature $T_{\mathrm{c}0}$ of the Ginzburg-Landau functional is reduced by Gaussian fluctuations according to Eq.~\eqref{shift2d}. 
Then, it is further reduced by vortex excitations according to Eq.~\eqref{shiftbkt}. 
We incorporated the effects of both amplitude fluctuations and phase fluctuations associated with vortex-antivortex excitations responsible for the BKT transition. 
With the semi-analytic relation between the bare superfluid phase stiffness and the renormalized one, we obtained a formula for the shift of the transition temperature including the topological vortex-antivortex excitations. 
Based on our approach to the cascade of fluctuations, we obtained the $H$-$T$ phase diagram of type-II superconductors and the critical behaviors of the heat capacity. 
The singular behavior of the heat capacity at criticality is hindered by phase fluctuations associated with the vortex-antivortex excitations, which is useful to verify which types of fluctuations dominate in a two-dimensional superconducting material. 

The Ginzburg-Landau theory is capable of describing multicomponent systems such as multiband superconductors as well as the single-component ones studied in this manuscript. 
In particular, the BKT transition is subject to change qualitatively due to the multicomponent character \cite{nitta2019,furutani2023,midei2023,midei2024,perali2023}. 
The clarification of the roles of fluctuations and their temperature evolution in the multicomponent two-dimensional superconductivity or superfluidity by extending this work to include multichannel fluctuations would be of great interest \cite{sala2019,kogan,grigorishin}.

\section*{Acknowledgments}

The authors thank A. A. Varlamov for the useful comments. 
This work has been partially supported by PNRR MUR project PE0000023-NQSTI. 
K.F. was supported by JSPS KAKENHI (Grant No.~JP24K22858) and Maki Makoto Foundation.

\appendix

\section{Determination of $\langle\abs{\eta}^{2}\rangle$}\label{Appeta}

\begin{figure}[t]
\centering
\includegraphics[keepaspectratio,scale=0.35]{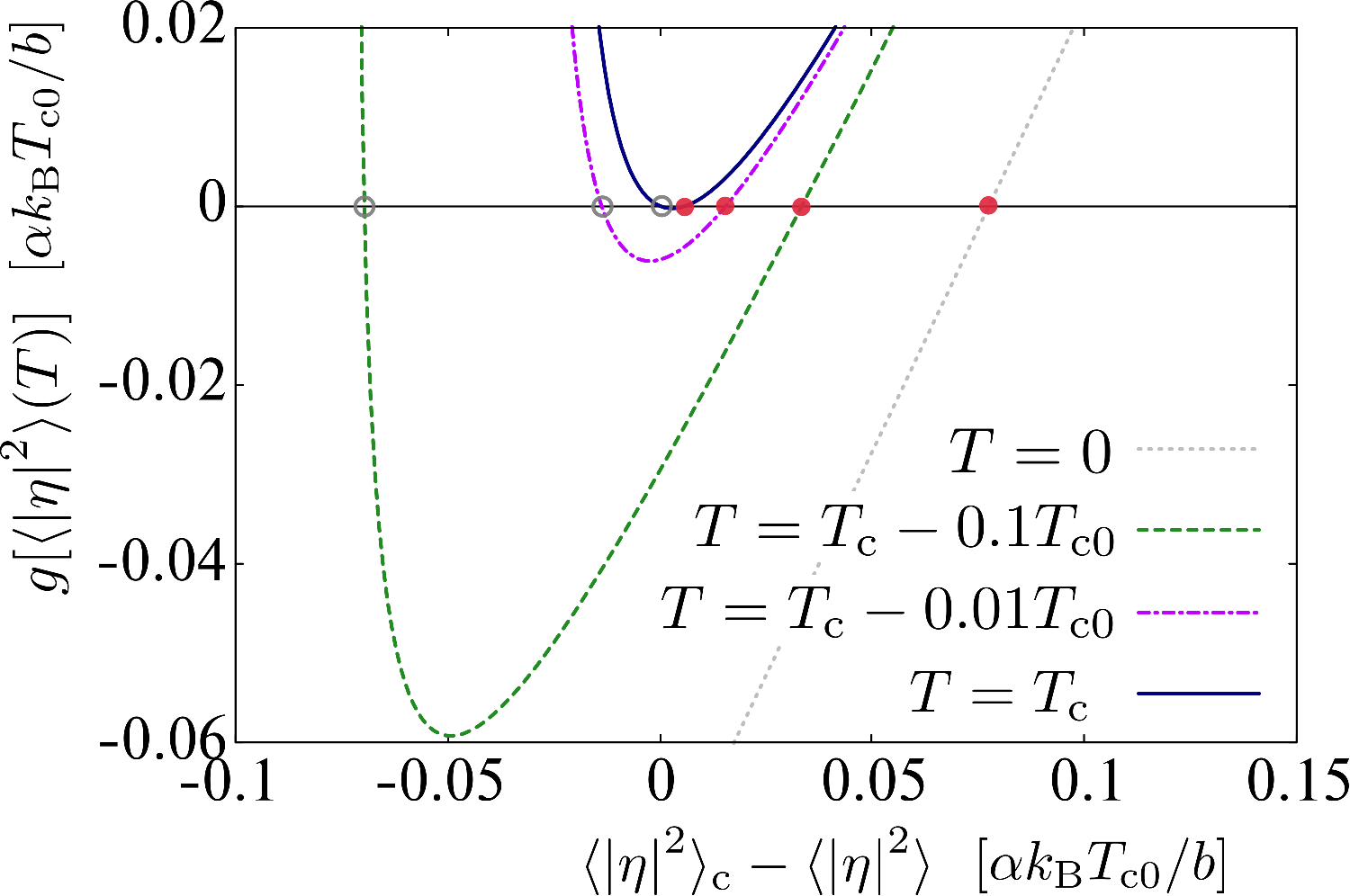}
\caption{Plot of $g[\langle\abs{\eta}^{2}\rangle]$ in Eq.~\eqref{geta} with respect to $\langle\abs{\eta}^{2}\rangle_{\rm c}-\langle\abs{\eta}^{2}\rangle$ below $T_{\rm c}$ with $\mathrm{Gi}=0.1$. 
The zero point corresponds to the root of Eq.~\eqref{eta2sf}. 
The filled red points satisfying $\langle\abs{\eta}^{2}\rangle_{\rm c}-\langle\abs{\eta}^{2}\rangle\ge0$ are physical roots, while the open circles in the region of $\langle\abs{\eta}^{2}\rangle_{\rm c}-\langle\abs{\eta}^{2}\rangle\le0$ are unphysical roots.} 
\label{Figetaroot}
\end{figure}

The self-consistent equation \eqref{eta2sf} determines $\langle\abs{\eta}^{2}\rangle$. 
Figure \ref{Figetaroot} shows a plot of 
\beq
\bal
&g[\langle\abs{\eta}^{2}\rangle(T)]\equiv\langle\abs{\eta}^{2}\rangle \\
&-\frac{k_{\rm B}T}{8\pi\gamma}\ln\left(\frac{1}{4\mathrm{Gi}}\frac{2\alpha k_{\rm B}(T_{\mathrm{c}0}-T)-4b\langle\abs{\eta}^{2}\rangle+\alpha k_{\rm B}T_{\rm c}/4}{2\alpha k_{\rm B}(T_{\mathrm{c}0}-T)-4b\langle\abs{\eta}^{2}\rangle+\mathrm{Gi}\alpha k_{\rm B}T_{\rm c}}\right), 
\eal
\label{geta}
\eeq
with $\mathrm{Gi}=0.1$. 
Equation \eqref{eta2sf} equivalent to $g[\langle\abs{\eta}^{2}\rangle(T)]=0$ determines $\langle\abs{\eta}^{2}\rangle$. 
There are two roots of Eq.~\eqref{eta2sf} for $\langle\abs{\eta}^{2}\rangle_{\rm c}-\langle\abs{\eta}^{2}\rangle\le0$ and $\langle\abs{\eta}^{2}\rangle_{\rm c}-\langle\abs{\eta}^{2}\rangle>0$ respectively with $\langle\abs{\eta}^{2}\rangle_{\rm c}\equiv \langle\abs{\eta}^{2}\rangle_{T\to T_{\rm c}^{+}}$ defined by Eq.~\eqref{eta2c}. 
The former is an unphysical root (open circles in Fig.~\ref{Figetaroot}) and the latter that vanishes at zero temperature is what we obtained to illustrate the quasiorder parameter in Fig.~\ref{FigTc0TcTBKT} (filled red points in Fig.~\ref{Figetaroot}). 
However, the self-consistent equation \eqref{geta} reveals that $\langle\abs{\eta}^{2}\rangle=\langle\abs{\eta}^{2}\rangle_{\rm c}$ at $T=T_{\rm c}$ is satisfied only by the unphysical root and the physical root still has a deviation $\langle\abs{\eta}^{2}\rangle_{T\to T_{\rm c}^{-}}>\langle\abs{\eta}^{2}\rangle_{\rm c}$ as shown in Fig.~\ref{Figetaroot}. 
This deviation of $\langle\abs{\eta}^{2}\rangle_{T\to T_{\rm c}^{-}}$ from $\langle\abs{\eta}^{2}\rangle_{\rm c}$ is diminished by a smaller $\mathrm{Gi}$ and vanishes in $\mathrm{Gi}\to0$. 
To consider the region in which the deviation is negligible, we consider $\mathrm{Gi}=0.1\ll1$ throughout this work. 

\section{Computation of heat capacity in $D$-dimensional superconductors and superfluids}\label{AppCv}

In this Appendix, we provide the calculation of the heat capacity considered in Sec.~\ref{Secheat} within the mean-field and Gaussian level with an extension to $D$-dimension. 
First of all, the contribution from the uniform superfluid order reads
\beq
\Delta C_{0}(T)=-\frac{T}{L^{D}}\pdv[2]{F_{\mathrm{s}0}(T)}{T},
\label{DeltaC0}
\eeq
with $F_{\mathrm{s}0}(T)=L^{D}\left[a(T)\psi_{0}^{2}+b\psi_{0}^{4}/2\right]$ at the MF level. 
At the GF and the BKT level, $F_{\mathrm{s}0}$ is replaced with Eqs.~\eqref{uniform-gf} and Eq.~\eqref{uniform-bkt}, respectively. 
Within the MF level, Eq.~\eqref{psi0} yields
\beq
\Delta C_{0}^{(\text{MF})}=
\begin{cases}
0 & (T\ge T_{\mathrm{c}0}), \\\\
\dfrac{\alpha^{2}k_{\rm B}^{2}}{b}T & (T< T_{\mathrm{c}0}).
\end{cases}
\eeq 
By including the thermal fluctuations within the GF level, Eqs.~\eqref{phi0} and \eqref{psiGFnearTc} give
\beq
\Delta C_{0}^{(\text{GF})}=
\begin{cases}
0 & (T\ge T_{\rm c}), \\\\
\dfrac{\alpha^{2}k_{\rm B}^{2}}{b}A_{-}(2-A_{-})T & (T\lesssim T_{\rm c}). 
\end{cases}
\eeq
The contribution from the fluctuations given by Eq.~\eqref{DeltaC} in the superfluid phase reads
\beq
\bal
\Delta C^{-}_{\rm fl}&=-\frac{T}{L^{D}}\pdv[2]{F_{\rm fl}^{-}(T)}{T} \\
&=-\dfrac{k_{\rm B}T}{L^{D}}\xi^{2}\sum_{\bf k}\dfrac{\partial_{T}\xi^{-2}+T\partial_{T}^{2}\xi^{-2}/2}{1+\xi^{2}k^{2}/2} \\
&+\dfrac{k_{\rm B}T^{2}}{2L^{D}}\xi^{4}\sum_{\bf k}\left(\dfrac{\partial_{T}\xi^{-2}}{1+\xi^{2}k^{2}/2}\right)^{2}, 
\eal
\label{DeltaC1}
\eeq
where the coherence length,
\beq
\xi(T)^{2}=
\begin{cases}
\dfrac{2\gamma}{a(T)+3b\psi_{0}^{2}} & (\text{MF}), \\\\
\dfrac{2\gamma}{a(T)+2b\langle\abs{\eta}^{2}\rangle+3b(\psi_{0}^{(\text{GF})})^{2}} & (\text{GF}), 
\end{cases}
\label{xi}
\eeq
coincides with Eq.~\eqref{xibelow} in the superfluid phase. 
In the vicinity of criticality, Eq.~\eqref{xi} yields the relations 
\bseq
\beq
\abs{\xi(T\to T_{\mathrm{c}0}^{-})^{2}}=\frac12\abs{\xi(T\to T_{\mathrm{c}0}^{+})^{2}} 
\eeq
\beq
\abs{\partial_{T}\xi(T\to T_{\mathrm{c}0}^{-})^{-2}}=2\abs{\partial_{T}\xi(T\to T_{\mathrm{c}0}^{+})^{-2}} 
\eeq
\label{xirelationMF}
\eseq
within the MF level, and
\bseq
\beq
\abs{\xi(T\to T_{\mathrm{c}}^{-})^{2}}=\frac{A_{-}}{2A_{+}}\abs{\xi(T\to T_{\mathrm{c}}^{+})^{2}} 
\eeq
\beq
\abs{\partial_{T}\xi(T\to T_{\mathrm{c}}^{-})^{-2}}=\frac{2A_{-}}{A_{+}}\abs{\partial_{T}\xi(T\to T_{\mathrm{c}}^{+})^{-2}} 
\eeq
\label{xirelationGF}
\eseq
by including the thermal fluctuations within the Gaussian level. 
Equation \eqref{DeltaC1} can be written as
\beq
\bal
&\Delta C^{-}_{\rm fl} \\
&=-\dfrac{\Omega_{D}k_{\rm B}T}{(2\pi)^{D}}\xi^{2-D}\left(\partial_{T}\xi^{-2}+\dfrac{T}{2}\partial_{T}^{2}\xi^{-2}\right)\int^{\xi\Lambda}_{\xi k_{0}}\dd{x}\dfrac{x^{D-1}}{1+x^{2}/2} \\
&+\dfrac{\Omega_{D}k_{\rm B}T^{2}}{2(2\pi)^{D}}\xi^{4-D}\left(\partial_{T} \xi^{-2}\right)^{2}\int^{\xi\Lambda}_{\xi k_{0}}\dd{x}\dfrac{x^{D-1}}{(1+x^{2}/2)^{2}}, 
\eal
\label{DeltaCsfD}
\eeq
where $\Omega_{D}=D\pi^{D/2}/\Gamma(D/2+1)$ is the volume of the $D$-dimensional unit sphere with $\Gamma(x)$ being the Gamma function. 
In the same manner, the heat capacity in the normal phase is given by
\beq
\bal
&\Delta C^{+}_{\rm fl}=-\frac{T}{L^{D}}\pdv[2]{F_{\rm fl}^{+}(T)}{T} \\
&=-\dfrac{2k_{\rm B}T}{L^{D}}\xi^{2}\sum_{\bf k}\dfrac{\partial_{T}\xi^{-2}+T\partial_{T}^{2}\xi^{-2}/2}{1+\xi^{2}k^{2}/2} \\
&+\dfrac{k_{\rm B}T^{2}}{L^{D}}\xi^{4}\sum_{\bf k}\left(\dfrac{\partial_{T}\xi^{-2}}{1+\xi^{2}k^{2}/2}\right)^{2} \\
&=-\dfrac{2\Omega_{D}k_{\rm B}T}{(2\pi)^{D}}\xi^{2-D}\left(\partial_{T}\xi^{-2}+\dfrac{T}{2}\partial_{T}^{2}\xi^{-2}\right)\int^{\xi\Lambda}_{\xi k_{0}}\dd{x}\dfrac{x^{D-1}}{1+x^{2}/2} \\
&+\dfrac{\Omega_{D}k_{\rm B}T^{2}}{(2\pi)^{D}}\xi^{4-D}\left(\partial_{T} \xi^{-2}\right)^{2}\int^{\xi\Lambda}_{\xi k_{0}}\dd{x}\dfrac{x^{D-1}}{(1+x^{2}/2)^{2}}.
\eal
\label{DeltaCnD}
\eeq
The most singular part is the second line in Eqs.~\eqref{DeltaCsfD} and \eqref{DeltaCnD} compared to each of the first terms, respectively. 
Using the relations \eqref{xirelationMF} and \eqref{xirelationGF}, the singular part of the heat capacity in the vicinity of criticality provides the relation 
\beq
\Delta C_{\rm fl,MF}^{-}=2^{D/2-1}\Delta C_{\rm fl,MF}^{+}, 
\label{DeltaCflMFD}
\eeq
within the MF level \cite{larkin}, and 
\beq
\Delta C^{-}_{\rm fl,GF}=2^{D/2-1}\left(\dfrac{A_{-}}{A_{+}}\right)^{4-D/2}\Delta C^{+}_{\rm fl,GF},
\label{DeltaCflGFD}
\eeq
within the Gaussian level if $\xi\Lambda\to+\infty$ and $\xi k_{0}\to 0$. 

For $D=2$, in particular, Eqs.~\eqref{DeltaCflMFD} and \eqref{DeltaCflGFD} recover Eqs.~\eqref{DeltaCflMFrelation} and \eqref{DeltaCflGFrelation}, respectively. 
Practically, one obtains the overall heat capacity as
\bseq
\beq
\bal
&\Delta C^{-}
=\Delta C_{0}+\Delta C_{\rm fl}^{-} \\
&=\Delta C_{0}-\frac{k_{\rm B}T}{4\pi}\left(\partial_{T}\xi^{-2}+\frac{T}{2}\partial_{T}^{2}\xi^{-2}\right)\ln{\left(\dfrac{1+\xi^{2}\Lambda^{2}/2}{1+\xi^{2}k_{0}^{2}/2}\right)} \\
&+\frac{k_{\rm B}T^{2}}{8\pi}\xi^{2}(\partial_{T}\xi^{-2})^{2}\left(\dfrac{1}{1+\xi^{2}k_{0}^{2}/2}-\dfrac{1}{1+\xi^{2}\Lambda^{2}/2}\right), 
\eal
\label{DeltaCsf2Dfull}
\eeq
\beq
\bal
&\Delta C^{+}=\Delta C_{\rm fl}^{+} \\
&=-\frac{k_{\rm B}T}{2\pi}\left(\partial_{T}\xi^{-2}+\frac{T}{2}\partial_{T}^{2}\xi^{-2}\right)\ln{\left(\dfrac{1+\xi^{2}\Lambda^{2}/2}{1+\xi^{2}k_{0}^{2}/2}\right)} \\
&+\frac{k_{\rm B}T^{2}}{4\pi}\xi^{2}(\partial_{T}\xi^{-2})^{2}\left(\dfrac{1}{1+\xi^{2}k_{0}^{2}/2}-\dfrac{1}{1+\xi^{2}\Lambda^{2}/2}\right). 
\eal
\label{DeltaCn2Dfull}
\eeq
\label{DeltaCsfn2Dfull}
\eseq
The fluctuation contributions in Eqs.~\eqref{DeltaCsfn2Dfull} imply that a smaller infrared momentum cutoff $k_{0}$ enhances the heat capacity.

\section{Heat capacity including vortex excitations in two-dimensional superconductors and superfluids}\label{AppCv2d}

The heat capacity including the vortex excitations can be obtained as
\beq
\Delta C_{\rm BKT}=\Delta C_{0}+\Delta C_{\mathrm{fl}},
\eeq
where $\Delta C_{0}$ is given by Eq.~\eqref{DeltaC0} with $D=2$ under the substitution of $\psi_{0}=\psi_{\rm R}^{(\text{BKT})}$ 
and $\Delta C_{\rm fl}\equiv \Delta C_{\mathrm{fl},\eta}+\Delta C_{\rm fl,XY}$ with
\beq
\bal
\Delta C_{\mathrm{fl},\eta}^{-} &=-\dfrac{T}{L^{2}}\pdv[2]{F_{\rm fl}^{-}}{T}\\
&=-\frac{k_{\rm B}T}{4\pi}\left(\partial_{T}\xi^{-2}+\frac{T}{2}\partial_{T}^{2}\xi^{-2}\right)\ln{\left(\dfrac{1+\xi^{2}\Lambda^{2}/2}{1+\xi^{2}k_{0}^{2}/2}\right)} \\
&+\frac{k_{\rm B}T^{2}}{8\pi}\xi^{2}(\partial_{T}\xi^{-2})^{2}\left(\dfrac{1}{1+\xi^{2}k_{0}^{2}/2}-\dfrac{1}{1+\xi^{2}\Lambda^{2}/2}\right),
\eal
\eeq
and
\beq
\xi(T)^{2}=\dfrac{2\gamma}{a(T)+2b\langle\abs{\eta}^{2}\rangle+\gamma\langle(\grad\Tilde{\theta})^{2}\rangle+3b(\psi_{\rm R}^{(\mathrm{BKT})})^{2}}.
\label{xiBKT}
\eeq
Moreover, the XY free energy in Eq.~\eqref{FflXY} also contributes to the heat capacity as
\beq
\bal
\Delta C_{\rm fl,XY}=-\dfrac{T}{L^{2}}\pdv[2]{F_{\rm fl,XY}}{T}. 
\eal
\eeq

\end{document}